\documentclass[%
 reprint,
10pt,
showpacs,preprintnumbers,
nofootinbib,
 amsmath,amssymb,
 aps,
 pra, 
]{revtex4-1}

\usepackage{graphicx}
\usepackage{dcolumn}
\usepackage{bm}
\usepackage{hyperref}
\usepackage{color}
\usepackage[normalem]{ulem}
\usepackage{amsmath}
\allowdisplaybreaks
\usepackage{dsfont}
\usepackage{adjustbox}
\usepackage{pgf,tikz}
\usepackage{pgfplots}
\usepackage{circuitikz}
\usepackage{siunitx}
\usepackage{hyperref} 
\graphicspath{{./}{Figures/}}

\newcommand{\inter}{\hat{\Lambda}_{\mathbf{n}}}

\begin{document}

\title{Efficient two-mode interferometers with spinor Bose-Einstein condensates}

\author{Artur Niezgoda, Dariusz Kajtoch and Emilia Witkowska}
\affiliation{Institute of Physics, PAS, Aleja Lotnik\'{o}w 32/46, PL-02-668 Warsaw, Poland}

\date{\today}

\begin{abstract}
We consider general three-mode interferometers using a spin-1 atomic Bose-Einstein condensate with macroscopic magnetization. 
We show that these interferometers, combined with the measurement of the number of particles in each output port, provide an ultra-high phase sensitivity.
We construct effective two-mode interferometers which involve two Zeeman modes showing that they also provide an
ultra-high phase sensitivity but of a bit reduced factor in the corresponding Fisher information.
A special case of zero magnetization is shown to persist the efficiency of the two-mode interferometry.
\end{abstract}


\maketitle

\section{Introduction}

Spinor Bose-Einstein condensates proved to be an ideal candidate for high resolution sensitive magnetometers~\cite{PhysRevLett.113.103004,PhysRevLett.98.200801,PhysRevA.88.031602,Kruger2005,Wildermuth2005,Wildermuth2006,Aigner1226}, where the information about magnetic field strength is encoded in $2F + 1$ magnetic sublevels of the total spin $F$ hyperfine manifold. Many theoretical proposals~\cite{PhysRevA.93.023627,PhysRevA.97.023616,PhysRevLett.111.180401,PhysRevA.93.033608,PhysRevLett.115.163002,PhysRevLett.118.150401} as well as experiments~\cite{PhysRevLett.117.013001,Lucke2011,Luo620,NaturePhys8,PhysRevLett.117.143004} have demonstrated generation of highly entangled quantum states in spinor condensates, thus opening up the possibility for entanglement-enhanced atomic magnetometers operating below the standard quantum limit (SQL). Exceptional functionality of atomic sensors is based on quantum interferometry techniques~\cite{RevModPhys.89.035002}. 

In the quantum interferometry scheme, a physical quantity like the magnetic field is mapped onto a phase difference $\theta$ between the internal states of the atoms, and can be estimated by performing a number of quantum measurements at the output~\cite{Smerzi}. The most widely utilized technique is based on the measurement of a single observable that provides a well behaved monotonic signal as a function of $\theta$. It may happen that in order to fully exploit the potential of the system a knowledge of the entire conditional probability distribution is necessary~\cite{Strobel2014,PhysRevLett.107.080504}. Regardless of the estimation strategy, the precision in the $\theta$ estimation is bounded from below via the Cram\'er-Rao inequality $\Delta \theta \geqslant 1/\sqrt{\nu \mathcal{I}}$~\cite{cramer}, where $\nu$ is the total number of measurements and $\mathcal{I}$ is the Fisher information (FI)~\cite{fisher_1925} which depends on the input state, interferometric protocol and measurement.
Clearly, modification of the measurement changes the estimation precision. According to the quantum Cram\'er-Rao theorem~\cite{Holevo2011} the FI cannot be larger than the quantum Fisher Information $F_Q$ (QFI), which is the  maximized FI over all allowable quantum measurements. Substituting $F_Q$ in place of $\mathcal{I}$ in equation for $\Delta \theta$ yields the ultimate lower bound on the precision achievable by a quantum mechanical strategy. The scaling of the QFI with the total atom number $N$ is of main interest. In the SQL, $F_Q \propto N$ is reached when uncorrelated atoms are used in the interferometry, while in the ultimate Heisenberg limit, $F_Q \propto N^2$ (HL) is possible by using entangled states. In principle, the precision in the magnetic field sensing can be significantly increased by employing entangled states in atomic magnetometers.

Here, we are interested in the interferometric utility of spin-1 Bose-Einstein condensates with three internal states numerated by the quantum magnetic number $m_F = 0, \pm 1$ for the situation of current experimental relevance \cite{RevModPhys.85.1191}, where the total number of atoms and the magnetization $M \equiv \langle \hat{J}_z \rangle =\langle \hat{N}_{m_F=1}\rangle - \langle\hat{N}_{m_F=-1} \rangle$ are both conserved. Additionally, the optical dipole trap is tight enough that the condensate forms in the single spatial mode with thermally populated internal degrees of freedom~\cite{PhysRevLett.119.050404}. In~\cite{PhysRevA.97.023616}, by calculating the QFI, we have shown that it is possible to overcome the SQL if the variance of magnetization $\Delta M$ is smaller than $\sqrt{N}$, and even approach the HL if $\Delta M<1$ when using such thermal states. In this work we pursue further our study focusing on an optimal measurement which maximizes the FI and on an accessible with current experimental techniques interferometric transformations.

In the paper we concentrate on the measurement of the number of particles in each Zeeman component showing that it maximizes the FI for the most optimal three-mode interferometric transformations.
Although, the three-mode interferometric transformations which optimally employ the QFI are not the representative ones, we show how they could be constructed in an experiment. 
We propose also to use a two-mode interferometric transformation equivalent to the Mach-Zehnder interferometer (MZI) in the appropriate choice of the SU(2) subspace. In practice, it means a reduction of the system to the two-mode description by truncation of the density matrix over the unused by the interferometer third mode. 
Then, by using such two-mode interferometric transformations the HL is still possible to reach with slightly reduced factor in some cases only.
We show also how to construct an effective two-mode interferometer for the system consisting of atoms having an arbitrary large value of the spin $F$.
The conclusion is suitable for any value of magnetization, including the special case of widely studied zero magnetization.
It is worth to notice that non-zero temperatures considered by us do not destroy the HL of the FI. Moreover, thermal fluctuations among internal degrees of freedom can be a resource for transition from the SQL to the HL in some cases.

In addition, we analyzed the estimation precision of effective two-mode interferometers using the error-propagation formula and measurement of the z-component of the total spin operator squared or parity operator. In general, the quantum Cram\'{e}r-Rao bound cannot be saturated in both cases, except $M = 0$ when the z-component of the total spin operator squared is optimal. However, in a special case when the third mode is not populated, the parity measurement is optimal for any value of magnetization.

Nonzero variance of magnetization, however, has destructive impact on the precision in the $\theta$ estimation as we have already pointed out in~\cite{PhysRevA.97.023616}. When $\Delta M >1$ the measurement of populations of Zeeman components is not the most optimal one as the FI slightly differs in the value from the QFI. However, the measurement is not the worst option because the FI shows, similarly as the QFI, that the SQL can be still overcome if the variance of magnetization is smaller than $\sqrt{N}$.

The paper is organized as follows. In Sections \ref{sec:model} and \ref{sec:QFI} we present the model and remind our previous results concerning the QFI values and optimal interferometric transformations. A two-mode interferometry is defined in Section \ref{sec:2mode}. An experimental implementation of optimal interferometric transformations is discussed in Section \ref{sec:implementation}. The main results of the paper are shown in Section \ref{sec:FI} for macroscopic magnetizations, and in Section \ref{sec:zeromagnetization} for zero magnetization. In last Section \ref{sec:moments} we examine the estimation precision from the measurement of signal based on the method of moments.

\section{The model}\label{sec:model}
The system we focus on is a spin-1 atomic condensate in a homogeneous magnetic field~\cite{KAWAGUCHI2012253,RevModPhys.85.1191}. We assume the single mode approximation is valid~\cite{Chang2005-bc,Zhang2005-ru,PhysRevA.60.1463,PhysRevA.66.011601,phdCorre}
\footnote{The system considered consists of a few thousand atoms in which creation of spin domains are energetically not favorable (spin healing length is much larger than the linear system size). Therefore, in the low temperature limit all atoms share the same spatial wave function and the spatial and spin degrees of freedom can be decoupled.},
and all atoms share the same spatial wave-function $\phi({\bf r})$, which is a solution of the Gross-Pitaevskii equation with normalization $\int d^3 r |\phi(\mathbf{r})|^2 = 1$~\cite{phdHamley}. The many-body system Hamiltonian reduces to~\cite{PhysRevA.82.031602,Luo620,PhysRevA.94.043623,Sarlo2013,PhysRevLett.111.180401,PhysRevA.93.033608,PhysRevA.92.023622,PhysRevA.97.023616}
\begin{equation}\label{eq:single_mode_ham}
\frac{\hat{\mathcal{H}}}{\tilde{c}} = \frac{{\rm sign}(c_2)}{2 N}\hat{J}^2 - q\hat{N}_{0},
\end{equation}
where $\hat{J}^2$ is the total spin operator and $\hat{N}_{m_F}$ is the particle number operator for the Zeeman state $m_F=0,\pm 1$. The energy unit is $\tilde{c} = N |c_2| \int d^3{r}|\phi(\mathbf{r})|^4$, where $c_2=4\pi\hbar^2(a_2 - a_0)/3\mu$,  $\mu$ is an atomic mass, and $a_0$ and $a_2$ are the s-wave scattering lengths~\cite{PhysRevA.66.011601}. For $c_2 < 0$ (e.g. rubidium-87) the interaction term favors the ferromagnetic phase, with maximal total spin length $J=N$, whereas for $c_2 > 0$ (e.g. sodium-23) the antiferromagnetic phase minimizes the interaction energy with spin length $J=0$~\cite{KAWAGUCHI2012253}. The second term in~\eqref{eq:single_mode_ham} describes the quadratic Zeeman energy, where $q=Q/\tilde{c}$ and $Q=(\mu_B {\cal B})^2/(4 E_{\rm hf})$
depends on the magnetic field strength ${\cal B}$, the Bohr magneton $\mu_B$ and the hyperfine energy splitting $E_{\rm hf}$ which can be both positive and negative \cite{PhysRevA.73.041602, TFScience2017}. The total number of atoms operator $\hat{N} = \sum_{m_F} \hat{N}_{m_F}$ and the z-component of the collective spin operator $\hat{J}_z =   \hat{N}_{+1}- \hat{N}_{-1}$ are both conserved. Thus, terms proportional to $\hat{N}$ and $\hat{J}_z$ have no influence on the results, and they were dropped in the final form of (\ref{eq:single_mode_ham}). The Hamiltonian~\eqref{eq:single_mode_ham} has a block-diagonal structure in the Fock state basis with each block labeled by the magnetization $M=-N,-N+1,\ldots,N$, being the eigenvalue of the $\hat{J}_z$ operator. 

Conservation of magnetization $M$ has a direct consequence on the equilibrium states of the spinor condensate~\cite{PhysRevLett.119.050404}. The general quantum state $\hat{\rho}$ takes the block-diagonal structure~\cite{Corre2015,PhysRevA.97.023616}
\begin{equation}\label{eq:rho_state}
\hat{\rho} = \sum\limits_{M=-N}^{N} w_{M}\hat{\rho}_{M},
\end{equation}
where $\hat{\rho}_{M} =\hat{P}_M e^{-\beta \hat{\mathcal{H}}/\tilde{c}} \hat{P}_M/\mathcal{Z}_{M}$ is a thermal state in the subspace of fixed magnetization $M$, with projection operator $\hat{P}_M$, and $\mathcal{Z}_{M}$ is the partition function ensuring $\text{Tr}\{\hat{\rho}_M\} = 1$. The temperature $T$ is controlled by the parameter $\beta = \tilde{c}/(k_B T)$, where $k_B$ is the Boltzmann constant. The non-thermal weights $w_M = \exp[-(M -\bar{M})^2/2\sigma^2]/Z$, where $Z = \sum_{M} \exp[-(M -\bar{M})^2/2\sigma^2]$, reflect experimental control over the magnetization before thermalization. The average value of magnetization is $\bar{M} = \langle \hat{J}_z\rangle$, while fluctuations of magnetization $\Delta M$ are set by $\Delta M = \sqrt{\langle \hat{J}_z^2\rangle - \langle \hat{J}_z\rangle^2} \simeq \sigma$.

\section{The Quantum Fisher Information}\label{sec:QFI}
The three magnetically sensitive Zeeman states can be used to encode information about unknown physical quantities using quantum interferometry techniques. In~\cite{PhysRevA.97.023616} the authors characterized metrological usefulness of quantum states defined in Eq.~\eqref{eq:rho_state} for generalized three-mode linear interferometry using the QFI. Whenever magnetization can be well controlled, $\sigma < 1$, both ground states and mixed by the temperature states provide Heisenberg-like scaling of the QFI, irrespective of the magnetization value. However, as noted in~\cite{PhysRevA.97.023616} fluctuations of magnetization reduce the QFI but the sub-SQL value preserves as long as $\sigma < \sqrt{N}$. 
The authors emphasize that the quantum interferometer which provides optimal value of the QFI is relatively simple due to rotational symmetry around the $\hat{J}_z$ operator of the state~\eqref{eq:rho_state}.

The output state of the three-mode interferometer with equal phase difference $\theta$ between neighboring paths can be written in general as
 $\hat{\rho}_{\rm out} =e^{-i \theta \hat{\Lambda}_{\mathbf{n}}} \hat{\rho} e^{i \theta \hat{\Lambda}_{\mathbf{n}}}$, where $\hat{\rho}$ is the input density matrix and $\hat{\Lambda}_{\mathbf{n}}=\mathbf{\hat{\Lambda}} \cdot \mathbf{n}$ is a generator of rotation, with a unit length vector $\mathbf{n}$ and a vector of generators $\mathbf{\hat{\Lambda}} = \{ \hat{J}_x, \hat{Q}_{zx}, \hat{J}_y, \hat{Q}_{yz}, \hat{D}_{xy}, \hat{Q}_{xy}, \hat{Y}, \hat{J}_z \} $, spanning the bosonic SU(3) Lie algebra:
 \begin{align}
 \hat{J}_{x} &\ =\ \frac{1}{\sqrt{2}}\left( \hat{a}^{\dag}_{\scriptscriptstyle{-1}}\hat{a}_{\scriptscriptstyle{0}} + \hat{a}^{\dag}_{\scriptscriptstyle{0}}\hat{a}_{\scriptscriptstyle{-1}} + \hat{a}^{\dag}_{\scriptscriptstyle{0}}\hat{a}_{\scriptscriptstyle{+1}} + \hat{a}^{\dag}_{\scriptscriptstyle{+1}}\hat{a}_{\scriptscriptstyle{0}}\right), \\
 \hat{Q}_{zx} &\ =\ \frac{1}{\sqrt{2}}\left( -\hat{a}^{\dag}_{\scriptscriptstyle{-1}}\hat{a}_{\scriptscriptstyle{0}} - \hat{a}^{\dag}_{\scriptscriptstyle{0}}\hat{a}_{\scriptscriptstyle{-1}} + \hat{a}^{\dag}_{\scriptscriptstyle{0}}\hat{a}_{\scriptscriptstyle{+1}} + \hat{a}^{\dag}_{\scriptscriptstyle{+1}}\hat{a}_{\scriptscriptstyle{0}}\right), \\
 \hat{J}_{y} &\ =\ \frac{i}{\sqrt{2}}\left( \hat{a}^{\dag}_{\scriptscriptstyle{-1}}\hat{a}_{\scriptscriptstyle{0}} - \hat{a}^{\dag}_{\scriptscriptstyle{0}}\hat{a}_{\scriptscriptstyle{-1}} + \hat{a}^{\dag}_{\scriptscriptstyle{0}}\hat{a}_{\scriptscriptstyle{+1}} - \hat{a}^{\dag}_{\scriptscriptstyle{+1}}\hat{a}_{\scriptscriptstyle{0}}\right), \\
 \hat{Q}_{yz} &\ =\ \frac{i}{\sqrt{2}}\left( -\hat{a}^{\dag}_{\scriptscriptstyle{-1}}\hat{a}_{\scriptscriptstyle{0}} + \hat{a}^{\dag}_{\scriptscriptstyle{0}}\hat{a}_{\scriptscriptstyle{-1}} + \hat{a}^{\dag}_{\scriptscriptstyle{0}}\hat{a}_{\scriptscriptstyle{+1}} - \hat{a}^{\dag}_{\scriptscriptstyle{+1}}\hat{a}_{\scriptscriptstyle{0}}\right), \\
 \hat{D}_{xy} &\ =\ \hat{a}^{\dag}_{\scriptscriptstyle{-1}}\hat{a}_{\scriptscriptstyle{+1}} + \hat{a}^{\dag}_{\scriptscriptstyle{+1}}\hat{a}_{\scriptscriptstyle{-1}}, \\
 \hat{Q}_{xy} &\ =\ i\left( \hat{a}^{\dag}_{\scriptscriptstyle{-1}}\hat{a}_{\scriptscriptstyle{+1}} - \hat{a}^{\dag}_{\scriptscriptstyle{+1}}\hat{a}_{\scriptscriptstyle{-1}}\right), \\
 \hat{Y} &\ =\ \frac{1}{\sqrt{3}}\left( \hat{a}^{\dag}_{\scriptscriptstyle{-1}}\hat{a}_{\scriptscriptstyle{-1}} - 2\hat{a}^{\dag}_{\scriptscriptstyle{0}}\hat{a}_{\scriptscriptstyle{0}} + \hat{a}^{\dag}_{\scriptscriptstyle{+1}}\hat{a}_{\scriptscriptstyle{+1}}\right),\\
 \hat{J}_{z} &\ =\  \hat{a}^{\dag}_{\scriptscriptstyle{+1}}\hat{a}_{\scriptscriptstyle{+1}} - \hat{a}^{\dag}_{\scriptscriptstyle{-1}}\hat{a}_{\scriptscriptstyle{-1}} ,
 \end{align}
 where $\hat{a}_{m_F}$ is the annihilation operator of the particle in the $m_F$ Zeeman component. For instance, the phase $\theta$ can be proportional to the magnetic field strength owing to the linear Zeeman effect~\cite{PhysRevA.97.023616}.

For macroscopic magnetization, i.e. $M=O(N)$ and $N-M = O(N)$, the optimal QFI was shown~\cite{PhysRevA.97.023616} to be $F_Q= 4{\rm max}(\lambda_A,\lambda_B)$, with
\begin{align}
\label{eq:eigenvalue1}
\lambda_A & =\Gamma_{66}, \\
\lambda_B & = \left( \Gamma_{33} + \Gamma_{44} + \sqrt{4\Gamma_{34}^2 + (\Gamma_{33} - \Gamma_{44})^2} \right)/2 
\label{eq:eigenvalue2},
\end{align}
where elements of the covariance matrix $\Gamma$ are defined as follows
\begin{align}\label{eq:covariance}
\Gamma_{i,j}[\hat{\rho}] & = \sum\limits_{k} v_k \left[ \frac{1}{2}\langle k
|\{\hat{\Lambda}_i,\hat{\Lambda}_j \} | k \rangle - \langle k|
\hat{\Lambda}_i |k \rangle \langle k| \hat{\Lambda}_j |k\rangle \right] \nonumber\\
& -
4\sum\limits_{k>l}\frac{v_k v_l}{v_k + v_l}\text{Re}\left[\langle k|
\hat{\Lambda}_i |l\rangle \langle k| \hat{\Lambda}_j |l\rangle\right],
\end{align}
with eigenvalues $v_k$ and eigenvectors $|k\rangle$ of the input density matrix operator $\hat{\rho} = \sum v_k |k\rangle\langle k|$. The maximal possible value of the QFI is $F_Q = 4N^2$ and sets the Heisenberg limit for the estimation precision $\Delta\theta$, which can be attained only by the fully particle entangled states. On the other hand, separable states can give at most $F_Q = 4N$. The factor $4$ in the scaling of characteristic limits of the QFI is due to the SU(3) Lie algebra (the extensions from qubits to qudits for corresponding scaling can be found in~\cite{PhysRevLett.96.010401,Smerzi}). 
In general, for separable qudit states $F_Q[\hat{\rho}_{sep}] \leqslant N(h_{\textrm{max}} - h_{\textrm{min}})^2$ ,
while for entangled states $F_{Q}[\hat{\rho}] \leqslant N^2 (h_{\textrm{max}} - h_{\textrm{min}})^2$, where
$h_{\textrm{max}}$ and $h_{\textrm{min}}$ are the maximal and minimal eigenvalues of the single qudit Hamiltonian, respectively. 
In the case of qutrits (three-mode case) one has $h_{\textrm{max}} = 1$ and $h_{\textrm{min}} = -1$, whereas for qubits (two-mode case) $h_{\textrm{max}} = 1/2$ and $h_{\textrm{min}} = -1/2$.

The optimal value of the QFI can be attained when particular interferometer, defined by the operator $\hat{\Lambda}_{\mathbf{n}}$, is used. When the QFI is determined by the value of $\lambda_A$ (or $\lambda_B$), the rotation $e^{-i \theta \hat{\Lambda}^{(A)}_{\mathbf{n}}}$ (or $e^{-i \theta \hat{\Lambda}^{(B)}_{\mathbf{n}}}$) optimizes the QFI, where
\begin{equation}
\label{eq:eigenvector1}
\hat{\Lambda}^{(A)}_{\mathbf{n}} = \hat{Q}_{xy}
\end{equation}
and
\begin{align}
\label{eq:eigenvector2}
&\hat{\Lambda}^{(B)}_{\mathbf{n}} = \frac{1}{\sqrt{\mathcal{N}}} (\hat{J}_y + \gamma \hat{Q}_{yz})\nonumber \\
&=
\frac{i(1-\gamma)}{\sqrt{2\mathcal{N}}} (\hat{a}^\dagger _{-1}\hat{a}_{0} - \hat{a}^\dagger _{0}\hat{a}_{-1}) 
 +
\frac{i(1+\gamma)}{\sqrt{2\mathcal{N}}} (\hat{a}^\dagger_{0}\hat{a}_{1} - \hat{a}^\dagger_{1}\hat{a}_{0}) 
\end{align}
with
\begin{equation}\label{eq:gamma}
\gamma =(\Gamma_{44}-\Gamma_{33}+\sqrt{4\Gamma_{34}^2 + (\Gamma_{33} - \Gamma_{44})^2})/(2\Gamma_{34})
\end{equation} 
and $\mathcal{N}=1 + \gamma ^2$. Although the covariance matrix~\eqref{eq:covariance} has a complicated form, the QFI is given by the variance of $\hat{\Lambda}^{(A,B)}$ as long as fluctuations of magnetization are negligible ($\sigma \to 0$). This property stems from the fact that generators of the rotation (\ref{eq:eigenvector1}-\ref{eq:eigenvector2}) change magnetization, and thus the second term in Eq.~\eqref{eq:covariance} vanishes. 

\renewcommand{\arraystretch}{1.5}
\begin{table}[ht]
\caption{The expressions for the quantum Fisher information $F_Q$ and the optimal generator of interferometer rotation $\hat{\Lambda}_{\mathbf{n}}$ when magnetization is macroscopic, i.e. $M=O(N)$ and $N-M = O(N)$.}
\label{tab:table1}
\begin{tabular}{c||c|c|c}\hline
& $\beta \to \infty$, $q<q_{\rm th}$ &  $\beta \to \infty$, $q>q_{\rm th}$ & $\beta \to 0$, any $q$ \\\hline\hline
$ F_Q $ & $4 \lambda_{A}$ & $4 \lambda_{B}$ & $4 \lambda_{B}$ \\ \hline
$ F_Q(\sigma \to 0) $ & $4 \langle ( \hat{\Lambda}^{(A)}_{\mathbf{n}})^2\rangle$ 
& $4 \langle ( \hat{\Lambda}^{(B)}_{\mathbf{n}})^2\rangle$ 
& $4 \langle ( \hat{\Lambda}^{(B)}_{\mathbf{n}})^2\rangle$ \\ \hline
$ \hat{\Lambda}_{\mathbf{n}}$ & $\hat{\Lambda}^{(A)}_{\mathbf{n}}$ & $\hat{\Lambda}^{(B)}_{\mathbf{n}}$ & $\hat{\Lambda}^{(B)}_{\mathbf{n}}$ \\ \hline
\end{tabular}
\end{table}

The diagram of the optimal QFI as a function of $q$ and $\beta$, for a fixed $M$, is very regular. It consists of three regions where the optimal interferometer is either $\hat{\Lambda}^{(A)}$ or $\hat{\Lambda}^{(B)}$ as summarized in Table~\ref{tab:table1}. In the zero temperature limit ($\beta \to \infty$), the optimal interferometer is $\inter^{(A)}$ for $q < q_{\textrm{th}}$, and changes into $\inter^{(B)}$ in the opposite case. The threshold value $q_{\textrm{th}}$ defines a degenerate point where both operators provide the same value of the QFI. Approximated formulas for the threshold point $q_{\rm th}$ are $q_{t-}\approx-1.2$ for $c_2<0$ and $q_{t+} \approx 0.8 m^2$ for $c_2>0$~\cite{PhysRevA.97.023616}. The zero-temperature behavior of the QFI extends deep into the non-zero temperature regime. When the thermal energy dominates, the QFI saturates at a finite value and does not depend on $q$, giving rise to a third region with optimal interferometer $\inter^{(B)}$. 

\section{The two-mode interferometer}\label{sec:2mode}
The optimal interferometer involves two modes $m_F=\pm 1$ only in the region A, what is not the case for B in general. However, when $|\gamma|=1$ the optimal interferometer in the region B is two-mode as well, e.g. $m_F=0,1$ for positive average magnetizations.
Therefore, we distinguish between the three-mode interferometer $\inter^{(B)}$, with proper value of $\gamma$ and the two-mode interferometer with
\begin{equation}
\hat{K}_{y} = i(\hat{a}^{\dagger}_{0}\hat{a}_{+1} - \hat{a}^{\dagger}_{+1}\hat{a}_{0}),
\end{equation}
where $\gamma=1$ was put into the in Eq.~\eqref{eq:eigenvector2} (as we concentrate on positive average magnetizations $\langle \hat{J}_z\rangle>0$).

\section{Experimental implementation of optimal interferometric rotations}\label{sec:implementation}
Magnetometers based on alkali atoms rely on the detection of Larmor precession. For a weak magnetic field oriented along the z-axis, the collective quantum state $\tilde{\rho}$ acquires the phase $\theta = \mathcal{S} \mathcal{B} t_{\textrm{hold}}$ during a hold time $t_{\textrm{hold}}$. The sensitivity $\mathcal{S}$ relates the Larmor frequency $\omega = \mathcal{S} \mathcal{B}$ to the magnetic field strength $\mathcal{B}$~\cite{Seltzer2008}. During the Larmor precession cycle, the state $\tilde{\rho}$ is rotated around the operator $\hat{J}_z$ according to the equation
\begin{equation}\label{eq:Larmor}
 \tilde{\rho}_\theta = e^{-i\theta\hat{J}_z} \tilde{\rho} e^{i\theta\hat{J}_z}.
\end{equation}
The states (\ref{eq:rho_state}) of spin-1 Bose-Einstein condensates are optimally employed if the interferometer is either $\inter^{(A)}$ or $\inter^{(B)}$. This can be done experimentally using current technology in a three stage interferometer with the phase imprinting process defined in Eq.~\eqref{eq:Larmor}. During a preparation procedure the input quantum state $\hat{\rho}$ is rotated using a unitary operator $\hat{R}$, resulting in $\tilde{\rho} = \hat{R} \hat{\rho} \hat{R}^{\dagger}$. Subsequently, the state $\tilde{\rho}$ is subject to the phase imprinting process~\eqref{eq:Larmor}. Finally, the state $\tilde{\rho}_\theta$ is dis-entangled using the conjugate rotation $\hat{R}^{\dagger}$ giving $\hat{\rho}_{\theta} = \hat{R}^{\dagger} \tilde{\rho}_\theta \hat{R}$. In order to implement the general rotation $\exp(-i\theta\inter)$ we need to find the unitary transformation $\hat{R}$ such that
\begin{equation}
\hat{R}^{\dagger} e^{-i\theta \hat{J}_z} \hat{R} = e^{-i\theta \inter}.
\end{equation}
It is a straightforward procedure for $\inter = \inter^{(A)} = \hat{Q}_{xy}$, because operators $\{\hat{D}_{xy}, \hat{Q}_{xy}, \hat{J}_z\}$ span the SU(2) Lie algebra, thus $\hat{R} = \exp(-i\pi\hat{D}_{xy}/4)$. In the second case, with $\inter = \inter^{(B)} = (\hat{J}_y + \gamma \hat{Q}_{yz})/\sqrt{1+\gamma^2}$, we need two rotations $\hat{R} = \exp(-i\pi\hat{J}_x/2) \exp(i\alpha \hat{Q}_{xy})$, where $\cos\alpha = (1+\gamma^2)^{-1/2}$. Both rotations can be realized experimentally, since they either involve two extremal modes $m_F = \pm 1$ or spin operators~\cite{PhysRevA.88.031602,doi:10.7566/JPSJ.82.094002}.

\section{The Fisher information}\label{sec:FI}
A large value of the QFI implies that the quantum state $\hat{\rho}$ may be useful for sub-SQL interferometry as long as a proper quantum measurement and an estimator are chosen. Finding the optimal measurement, which at the same time is experimentally implementable, is not a straightforward task. A collection of measurement operators $\hat{\Pi}_x$, satisfying $\sum_x \hat{\Pi}_x^{\dagger}\hat{\Pi}_x = \mathds{1}$, defines conditional probabilities $p(x|\theta) = \text{Tr}\{\hat{\Pi}_x\hat{\rho}_{\rm out}\}$ of measuring the outcome $x$ given the $\theta$. The knowledge of $p(x|\theta)$ is used to construct an estimator for the phase $\theta$, and according to the Cram\'{e}r-Rao inequality the precision $\Delta \theta$ in the $\theta$ estimation is bounded from below by the Fisher information $\mathcal{I}(\theta)$ defined as~\cite{Pezze2014,fisher_1925}
\begin{equation}\label{eq:fisher_definition}
\mathcal{I}(\theta) = \sum\limits_{x} \frac{1}{p(x|\theta)}\left(\frac{\partial p(x|\theta)}{\partial\theta}\right)^2.
\end{equation}
An equivalent formula for the FI can be found in Appendix~\ref{ap:FI}.
The FI is always smaller than the QFI, i.e. $\mathcal{I}(\theta) \leqslant F_Q$, because the QFI is already optimized over all possible quantum measurements~\cite{Holevo2011}. Suggesting a measurement $\hat{\Pi}_x$ that would saturate this inequality is a relevant theoretical task for practical implementation. As noted in~\cite{PhysRevLett.72.3439,Nagaoka2005}, there always exists a projective measurement strategy in the eigenbasis of the symmetric logarithmic derivative~\cite{Holevo2011,Pezze2014} that is optimal, but it is $\theta$-dependent and typically not straightforward. 

Our choice is the experimentally relevant measurement of populations of Zeeman components, and the purpose of our calculations is to verify under which circumstances this measurement is optimal, i.e. maximizes the FI up to its quantum value.
We define the operator of the measurement as $\hat{\Pi}_{\{M,k\}}=|M,k \rangle \langle M,k|$, where $|M,k \rangle \equiv |N_{+1}, N_0, N_{-1}\rangle$ is the Fock state with constraint imposed on the total particle number $\sum_{m_F} N_{m_F} = N$, the magnetization $N_{+1} - N_{-1} = M$ and $k = N_{+1}$. The occupation of the $m_F=+1$ Zeeman component depends on the magnetization $k=k_{\textrm{min}}, \ldots, k_{\textrm{max}}$, where $k_{\textrm{min}} = \max(0,M)$ and $k_{\textrm{max}} = \lfloor (N+M)/2 \rfloor$. 
The probability distribution $p(\{M,k\}|\theta)$ is then
\begin{equation}\label{eq:probabilitydistribution}
 p(\{M,k\}|\theta) = 
 \langle M,k| e^{-i \theta \hat{\Lambda}_{\mathbf{n}}} \hat{\rho} e^{i \theta \hat{\Lambda}_{\mathbf{n}}}  | M,k\rangle.
\end{equation}
In the parametrized Fock state basis, the general quantum states \eqref{eq:rho_state} can be written as
\begin{equation}
 \hat{\rho} = 
 \sum\limits_{M'}w_{M'} \sum\limits_{k_1,k_2} \rho^{M'}_{k_1,k_2}|M',k_1\rangle\langle M',k_2|
\end{equation}
and the probability distribution~\eqref{eq:probabilitydistribution} is equal to
\begin{align}
 & p(\{M,k\}|\theta) =\nonumber \\
 & \sum\limits_{M'} w_{M'} \sum\limits_{k_1,k_2} \rho^{M'}_{k_1,k_2} \left[\mathcal{D}^{M',M}_{k_1,k}(\hat{\Lambda}_{\mathbf{n}},\theta)\right]^{*}  \mathcal{D}^{M',M}_{k_2,k}(\hat{\Lambda}_{\mathbf{n}},\theta),
\end{align}
where $\mathcal{D}^{M',M}_{k',k}(\hat{\Lambda}_{\mathbf{n}},\theta) = \langle M',k'| e^{i\theta\hat{\Lambda}_{\mathbf{n}}}|M,k\rangle$. 

Below we will calculate $\mathcal{I}(\theta)$ analytically and numerically based on (\ref{eq:fisher_definition}) for the ground and thermal states (\ref{eq:rho_state}) of the system (\ref{eq:single_mode_ham}) with fixed and fluctuating magnetization.

\subsection{Ground states and fixed magnetization}
We start our analysis with the simplest case of the pure state $|\Psi_{\bar{M}}\rangle$ in the block of fixed magnetization $\bar{M}$. It means that the distribution of magnetization is just $w_M=\delta_{\bar{M}, M}$. We denote the rotated pure state by $|\Psi_{\bar{M}}\rangle_\theta \equiv e^{-i \theta \hat{\Lambda}_{\mathbf{n}}}|\Psi_{\bar{M}}\rangle$ which expanded over the Fock state basis is $|\Psi_{\bar{M}}\rangle_\theta =\sum_{M,k} C_{M,k}(\theta) |M,k\rangle$. Expansion coefficients $C_{M,k}(\theta)$ are all real, because both the state $|\Psi_{\bar{M}}\rangle$ and the rotations with $\hat{\Lambda}_{\mathbf{n}}^{(A,B)}$ operators are real. The conditional probability is equal to $p(\{M,k\}|\theta) = C_{M,k}^2(\theta)$. Since the probability is determined by a single real number, one can show that the FI equals to~\cite{Wasak2016}
 \begin{equation}\label{eq:FI_pure_state}
  \mathcal{I}(\theta) = 4\langle \Psi_{\bar{M}}| \hat{\Lambda}_{\bf n}^2 | \Psi_{\bar{M}} \rangle,
 \end{equation}
and does not depend on $\theta$. The FI~\eqref{eq:FI_pure_state} takes the same form as the QFI~\cite{PhysRevA.97.023616}. It holds for both the three- and two-mode interferometers. We mentioned that operators $\hat{\Lambda}_{\mathbf{n}}^{(A,B)}$ give the optimal value of the QFI for macroscopic magnetization, but in fact they are also good in the case $\bar{M} = 0$. The slight difference is that for $c_2 < 0$ the optimal operator can be $\hat{J}_y$ ($\gamma = 0$ in $\hat{\Lambda}_{\mathbf{n}}^{(B)}$) for some values of magnetic field, see Section~\ref{sec:zeromagnetization}.

\begin{figure}[ht]
\begin{picture}(0,200)
\put(-130,100){\includegraphics[width=0.5\linewidth]{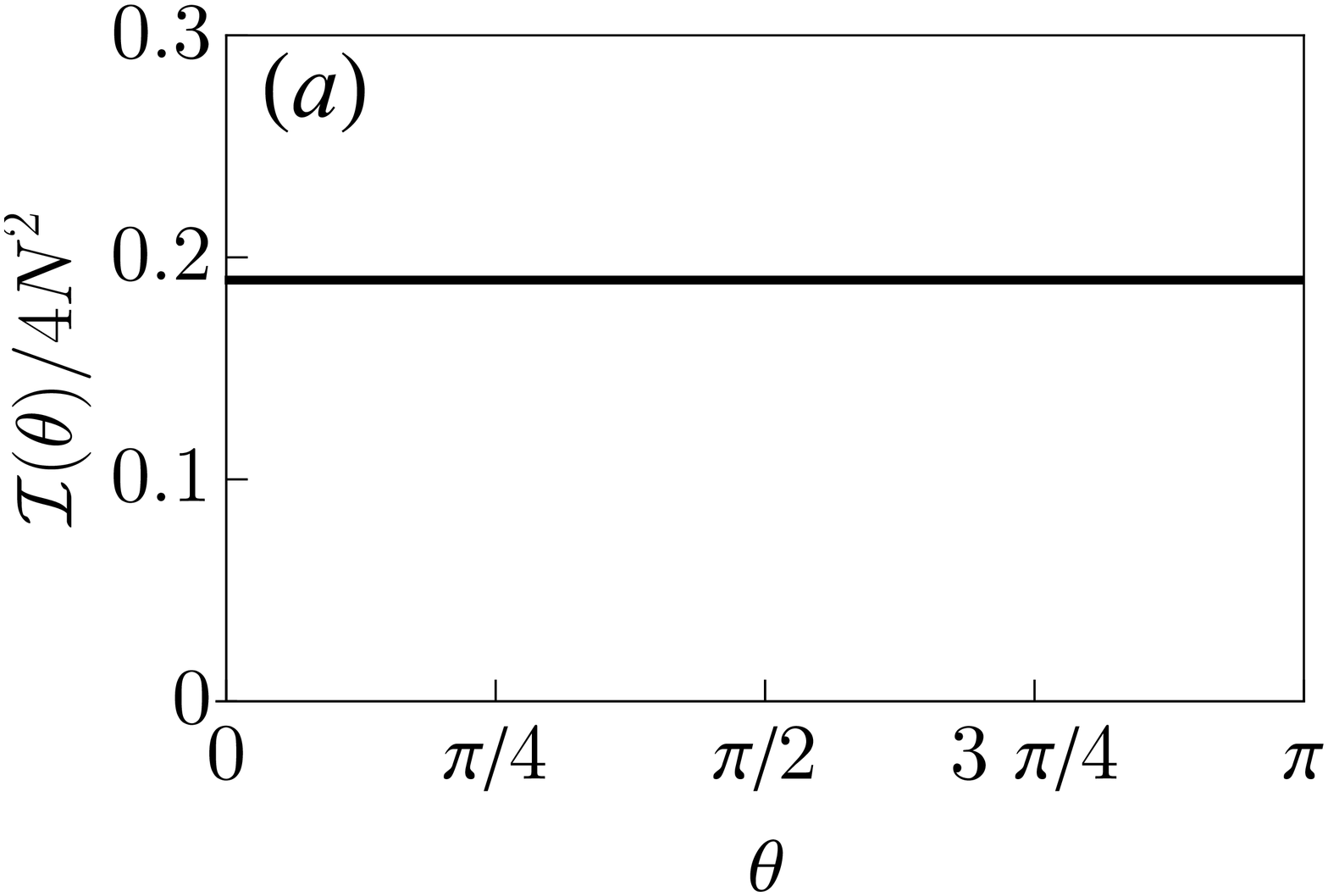}}
\put(-0,100){\includegraphics[width=0.5\linewidth]{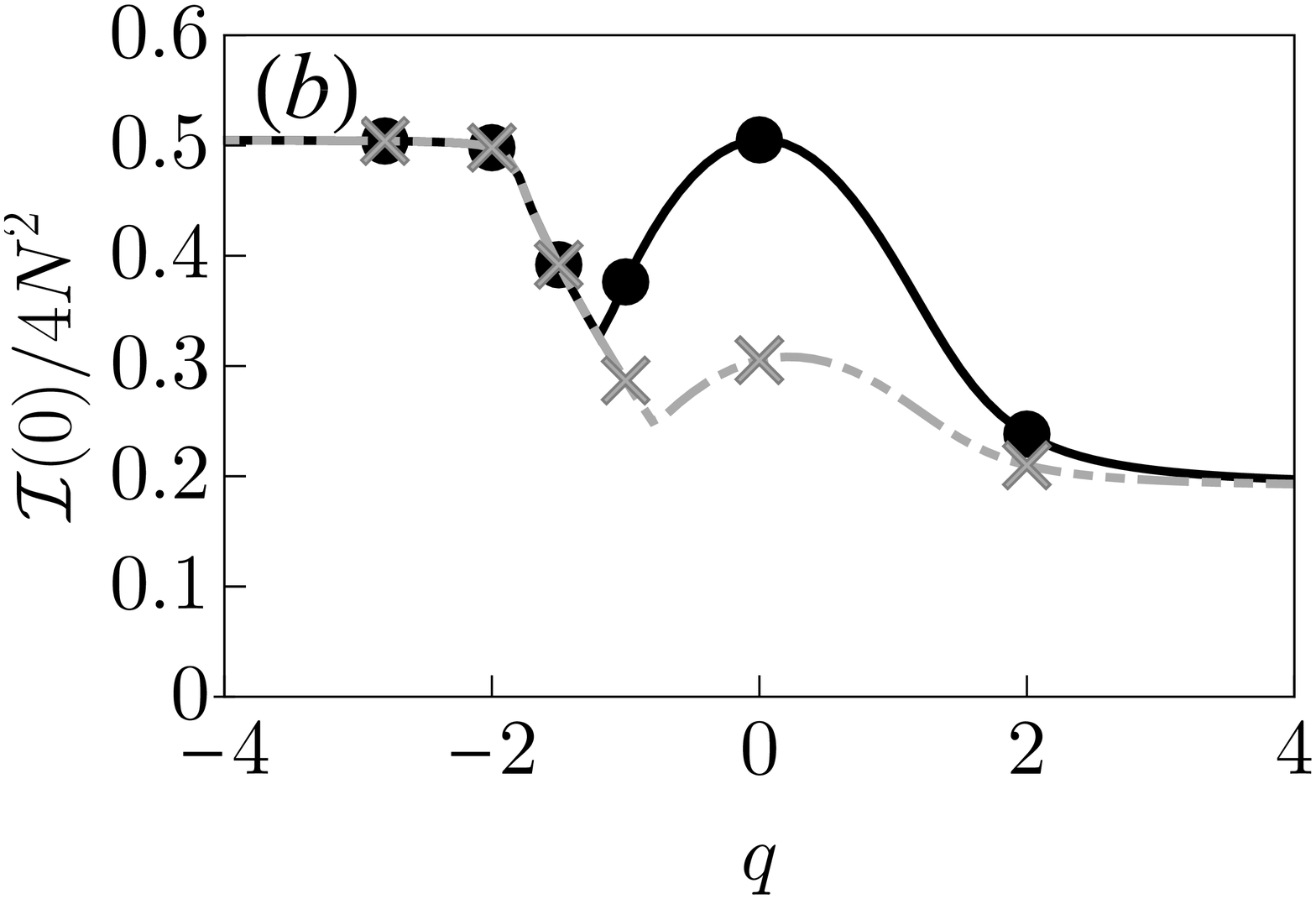}}
\put(-130,0){\includegraphics[width=0.5\linewidth]{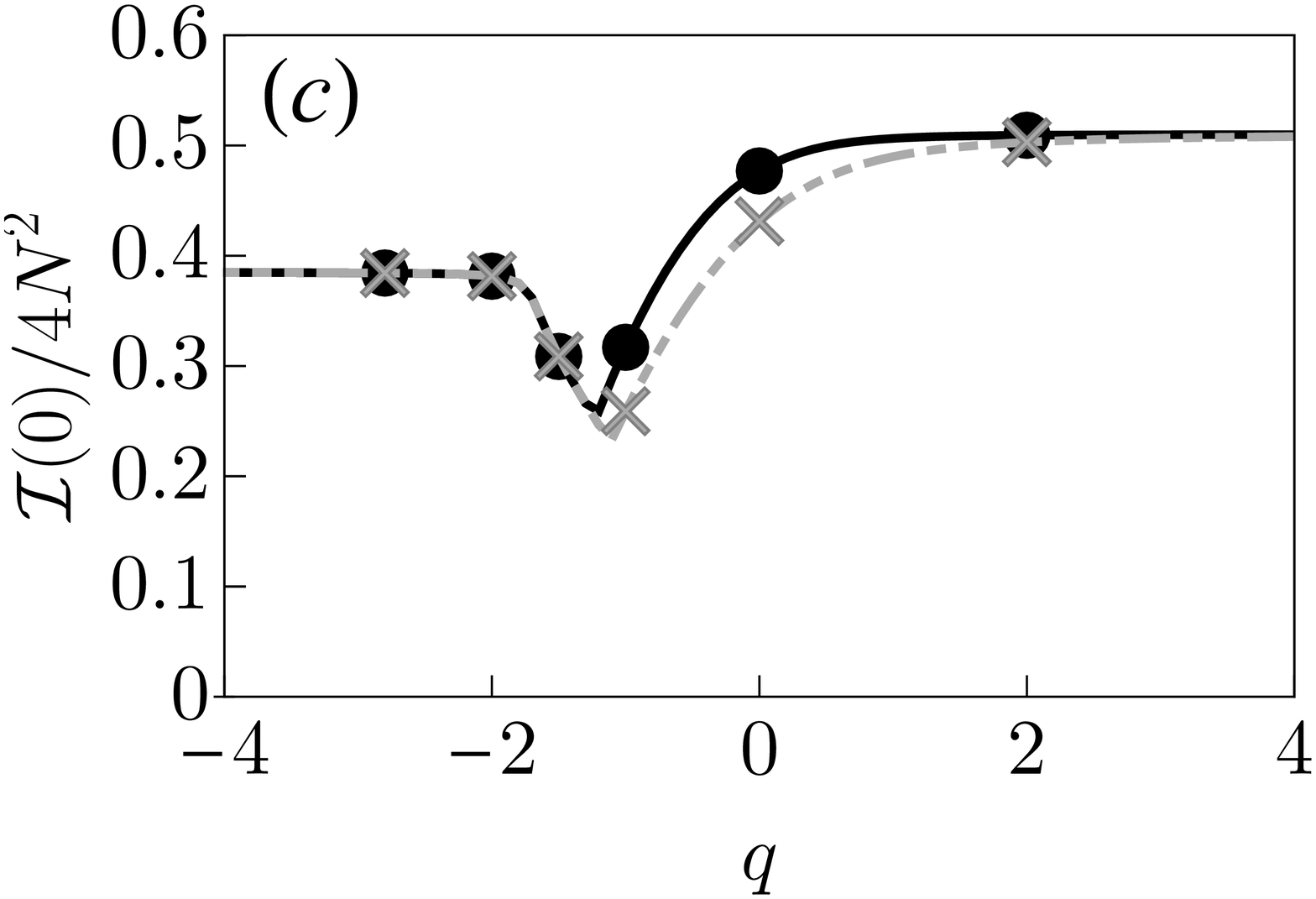}}
\put(-0,0){\includegraphics[width=0.5\linewidth]{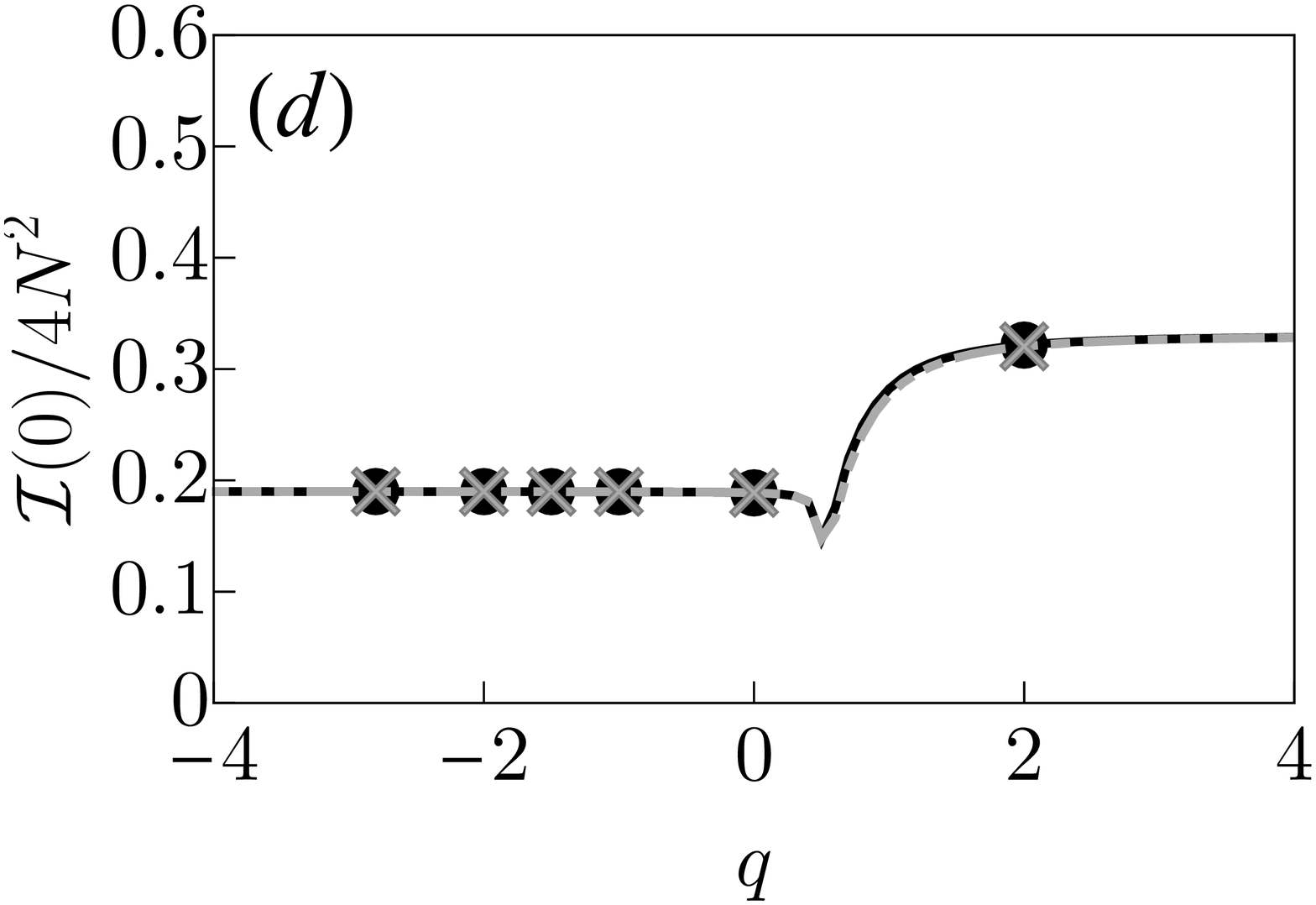}}
\end{picture}
\caption{(Color online) An example of the FI~(\ref{eq:fisher_definition}) divided by $4N^2$ versus $\theta$ from exact numerical calculations for $M=0.8N$, $q=0$ and $c_2>0$ is shown in $(a)$. The FI at $\theta=0$ versus $q$ for the three-mode (black circles) and two-mode (gray crosses) interferometry compared to the QFI for the three-mode (black solid lines) and two-mode (gray dashed lines) interferometry in the ground state of the system for fixed magnetization $M=0.1N$ and $c_2<0$ in $(b)$, $M=0.5N$ and $c_2<0$ in $(c)$, and $M=0.8N$ and $c_2>0$ in $(d)$. The total atom number is $N=10^2$.}
\label{fig:fig1}
\end{figure}

We verified our finding by numerical calculations within the exact diagonalization method, see Appendix~\ref{ap:numerics} for explanation. An example of numerical results is shown in Fig.\ref{fig:fig1} demonstrating validity of the analytical analysis.

\subsection{Thermal states and fixed magnetization}
When the temperature is non-zero, but the state still has a well defined magnetization ($\sigma \to 0$), the general quantum state~\eqref{eq:rho_state} can be written down in the Fock state basis as follows
\begin{equation}
 \hat{\rho} = \sum\limits_{k_1,k_2} \rho^{\bar{M}}_{k_1,k_2}|\bar{M},k_1\rangle \langle \bar{M},k_2|.
\end{equation}
When the information about $\theta$ is encoded on three modes, e.g. by rotation around $\hat{\Lambda}^{(B)}_{\mathbf{n}}$, analytical expressions for the FI can be derived only at $\theta = 0$. 

The probability $p(\{M,k\}|0) = \rho^{\bar{M}}_{k_1,k_1}\delta_{\bar{M},M}\delta_{k_1,k}$ is non-zero only for limited values of $\{M,k\}$. On the other hand, the first derivative 
\begin{align}
&\left. \frac{\partial p(\{M,k\}|\theta)}{\partial \theta}\right|_{\theta=0} = 
\sum\limits_{k_1,k_2} \rho^{\bar{M}}_{k_1,k_2}  \delta_{\bar{M},M} \, \, \cdot \nonumber \\
&\cdot \left\{
\langle \bar{M},k_1| (i\hat{\Lambda}_{\bf n}) |M,k \rangle^{*} \delta_{k_2,k}
 +  \langle M,k_2|(i\hat{\Lambda}_{\bf n})|\bar{M},k\rangle \delta_{k_1,k} \right\}
\end{align}
is always zero for operators $\hat{\Lambda}_{\mathbf{n}}^{(A,B)}$, because they do change magnetization. Whenever $p(\{M,k\}|0)\to 0$ one has to analyze the $0/0$ expression, e.q. using l'Hospital's rule. In that case, the FI is determined by the second derivative of the probability $p(\{M,k\}|\theta)$ taken at $\theta = 0$, which gives
\begin{align}
\mathcal{I}(0) = &4 \sum\limits_{k_1,k_2} \rho^{\bar{M}}_{k_1,k_2} \langle \bar{M},k_2|\hat{\Lambda}_{\bf n}^2|\bar{M},k_1\rangle  \nonumber \\
 -& 4\sum\limits_{k}\sum\limits_{k_1,k_2} \rho^{\bar{M}}_{k_1,k_2}\langle \bar{M},k|\hat{\Lambda}_{\bf n}|\bar{M},k_1\rangle\langle \bar{M},k_2|\hat{\Lambda}_{\bf n}|\bar{M},k\rangle,\label{eq:fisherclasstermal}
\end{align}
where the summation runs over all indexes for which $\rho^{M}_{k,k} \neq 0$. The operators $\hat{\Lambda}_{\mathbf{n}}$ that do not change the magnetization, e.q. $\hat{\Lambda}_{\mathbf{n}}=\hat{J}_z$ or $\hat{Y}$, give $\mathcal{I}(0) = 0$ due to compensation of both terms in Eq.~\eqref{eq:fisherclasstermal}. Luckily, for operators $\hat{\Lambda}^{(A,B)}_{\mathbf{n}}$ the second term in~\eqref{eq:fisherclasstermal} vanishes and the FI equals the QFI, see Table~\ref{tab:table1}. However, the $\theta$ dependence of the FI is in general unknown, and therefore numerical calculations are needed.

\begin{figure}[ht]
\begin{picture}(0,180)
\put(-130,95){\includegraphics[width=0.5\linewidth]{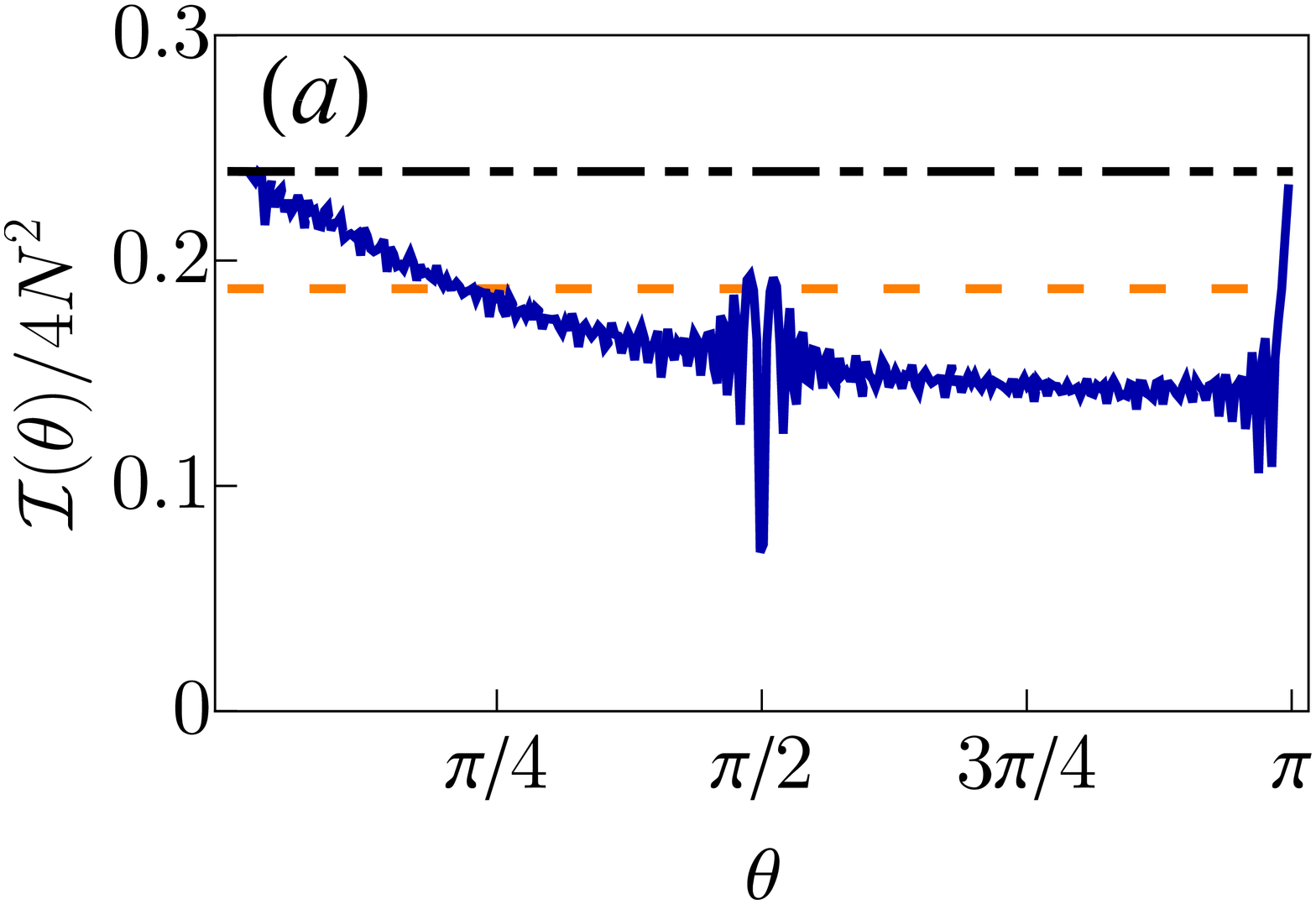}}
\put(-0,95){\includegraphics[width=0.5\linewidth]{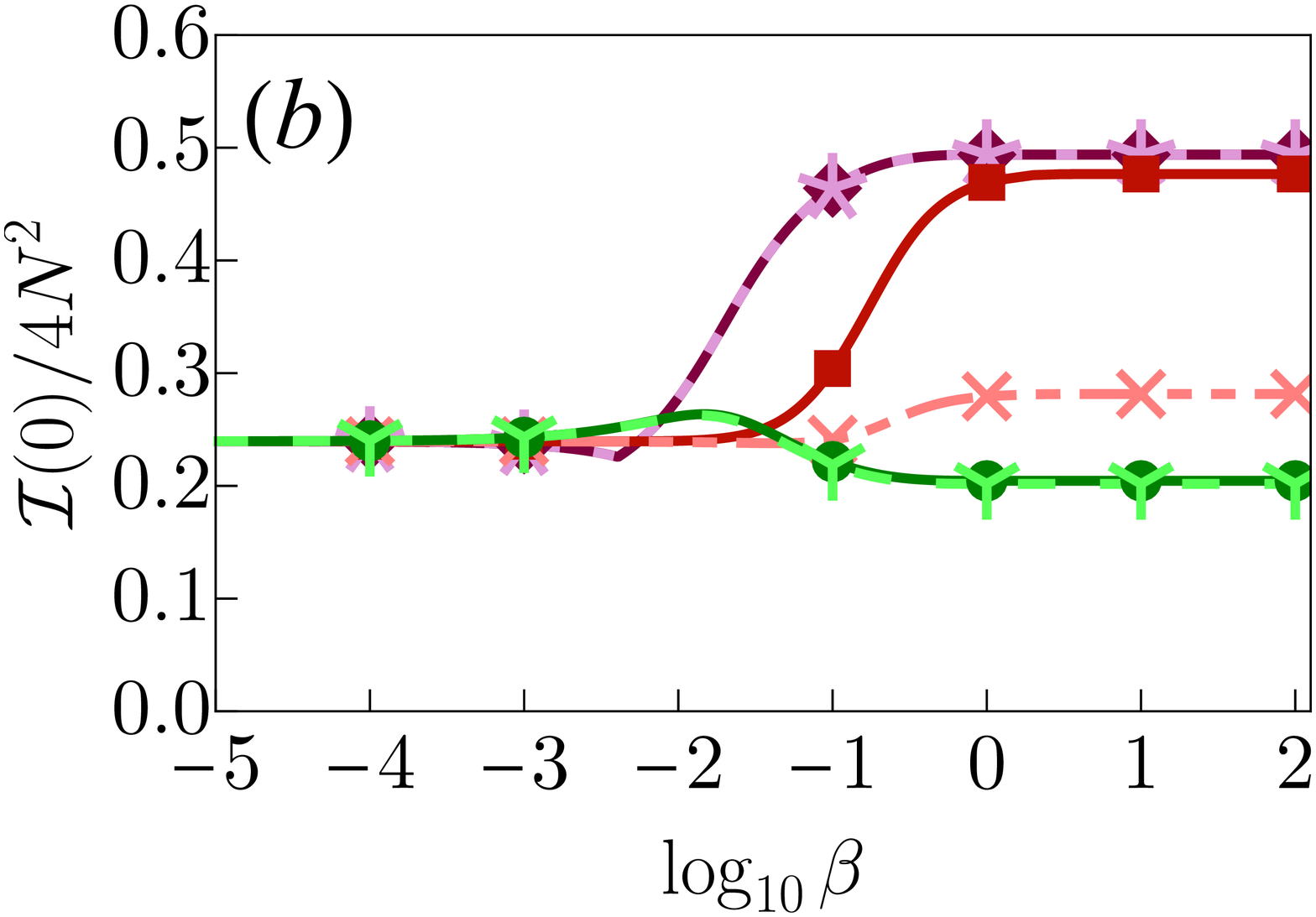}}
\put(-130,0){\includegraphics[width=0.5\linewidth]{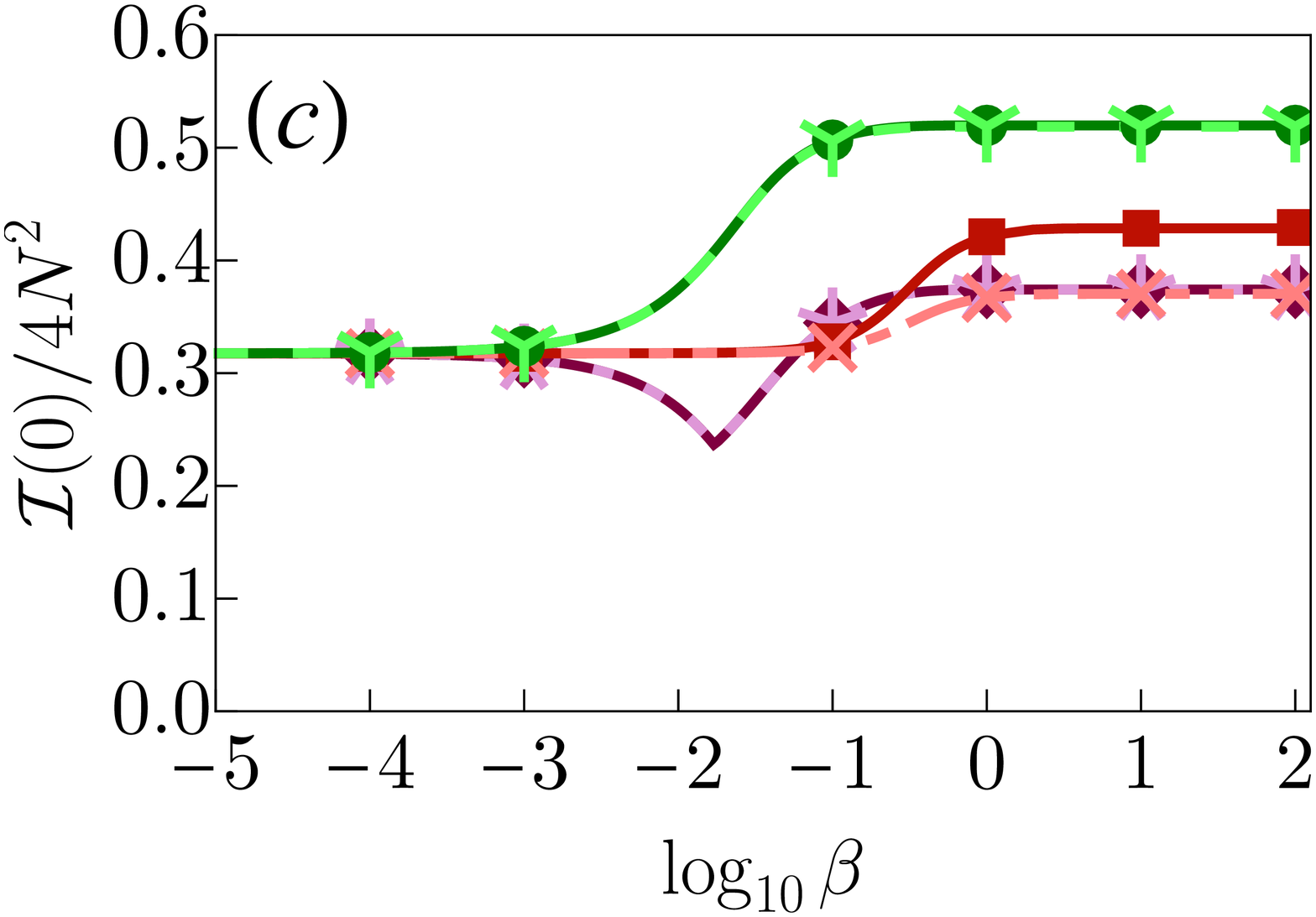}}
\put(-0,0){\includegraphics[width=0.5\linewidth]{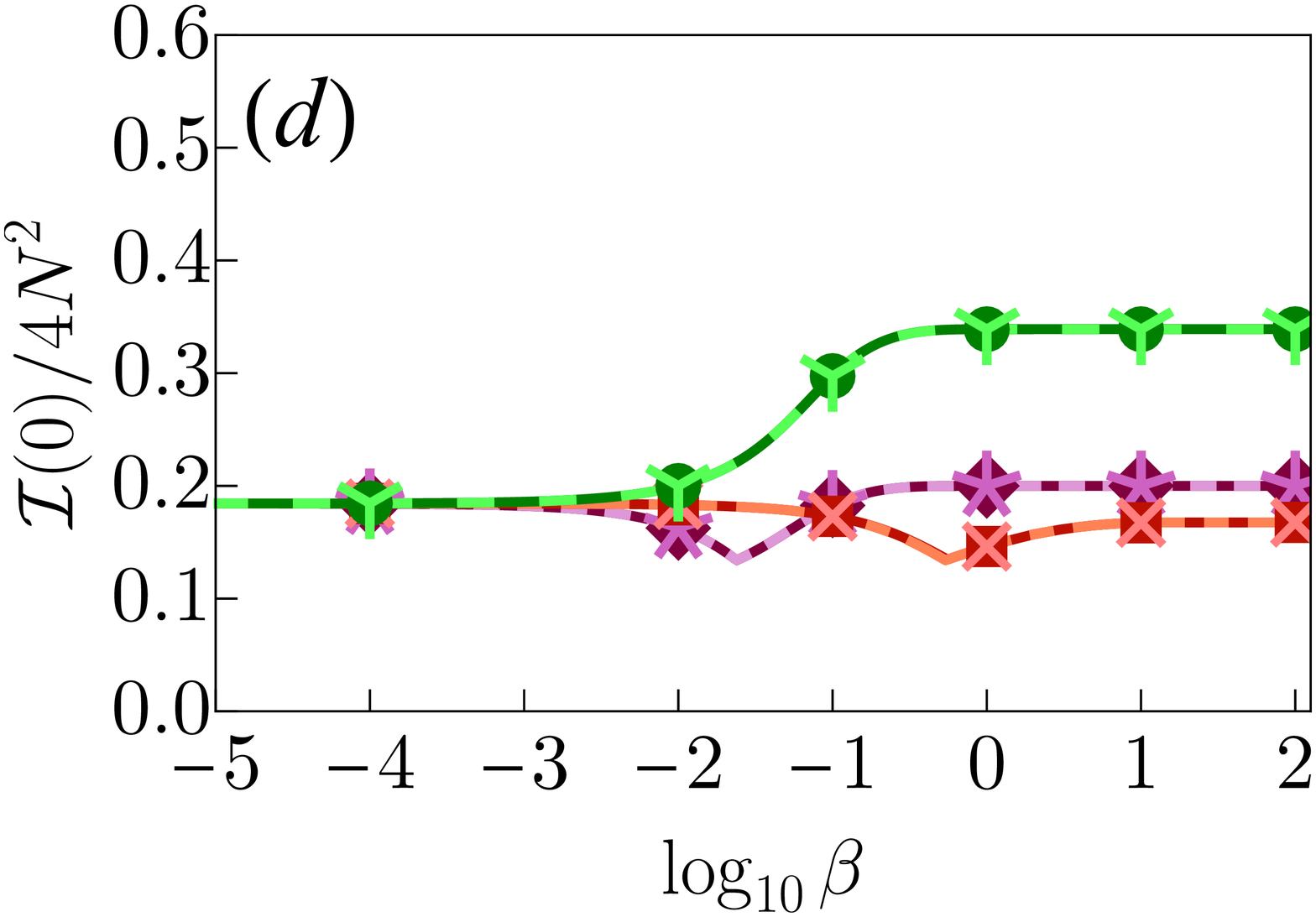}}
\end{picture}
\caption{(Color online) $(a)$ An example of $\theta$ dependence of the FI~\eqref{eq:fisher_definition} divided by $4N^2$ for  $\hat{\Lambda}_{\bf n}=\hat{\Lambda}^{(A)}_{\bf n}$ (orange dashed line), the two-mode $\hat{\Lambda}_{\bf n}=\hat{K}_{y}$ (black dash-dotted line) and the three-mode $\hat{\Lambda}_{\bf n}=\hat{\Lambda}^{(B)}_{\bf n}$ (blue solid line) interferometric transformations with $M=0.1N$, $\beta = 10^{-3}$, $q=-0.5$, $c_2<0$.
In $(b)$-$(d)$ the two-mode (crossed points) and the three-mode (filled points) versions of the FI divided by $4N^2$ versus $\log_{10}\beta$ are shown and compared to the two-mode (dashed lines) and three-mode (solid lines) versions of the QFI divided by $4N^2$ for different values of $q$ as indicated by colors: $q=-5$ (marked by purple), $q=-0.5$ if $c_2<0$ or $q=0.5$ if $c_2>0$ (marked by red) and  $q=5$ (marked by green). The parameters are $M=0.1N$, $c_2<0$ in $(b)$, $M=0.5N$, $c_2<0$ in $(c)$, $M=0.8N$, $c_2>0$ in $(d)$ with $N=50$.}
\label{fig:fig2}
\end{figure}

By performing and analyzing exact numerical results we made the following observations. In general, the FI for the three-mode interferometer depends on the phase $\theta$, taking the maximal value for $\theta = 0,\pm \pi, \dots$. 
We noticed that variation of $\mathcal{I}(\theta)$ versus $\theta$ strongly depends on the value of $q$. In the vicinity of the threshold point $q_{\textrm{th}}$ the FI rapidly varies with $\theta$ (as demonstrated in Fig.~\ref{fig:fig2}$(a)$). The farther away from the $q_{\textrm{th}}$, the smaller the changes, which eventually gives $\mathcal{I}(\theta)\to \mathcal{I}(0)$ when $|q|\gg q_{\textrm{th}}$. 
In the high magnetic field limit, $q \to \infty$, the FI for the two-mode transformations are equal to the FI for the three-mode interferometers, and they are $\theta$ independent. 
This observation suggests that the two-mode interferometers may in general give the FI independent of $\theta$. Indeed, we confirmed both numerically (see Fig.~\ref{fig:fig2}) and analytically (see Section~\ref{sub:mappingsu2}) that the FI and the QFI have the same value for the two-mode interferometers and they are independent of $\theta$. We explain that surprising at the first look result for our system in Section \ref{sub:mappingsu2} and generalize to the systems composed of atoms with higher spin in Section \ref{sub:mappingsuany}. 
In Fig.~\ref{fig:fig2} we gathered exact numerical results for a broad range of parameters. Comparison of the QFI value to the maximal value of FI is shown in panels $(b)$-$(d)$.
The overlap between the two-mode (crossed points) and three-mode (closed points) versions of the FI, and the two-mode (dashed lines) and three-mode (solid lines) versions of the QFI is clearly visible. It demonstrates that the two-mode interferometer transformation is competitive with the three-mode one, and gives the HL of the FI in a broad range of parameters.

\subsubsection{Mapping to the SU(2)}\label{sub:mappingsu2}
When the unknown value of $\theta$ is encoded only in two modes, the third mode carries no information about $\theta$, and any measurement involving that mode does not increase the FI value. The interferometric precision is the same as for the initial quantum state $\hat{\rho}$ traced out over the unused mode $\hat{\rho}_{2} = \text{Tr}_3\{\hat{\rho}\}$. This is a general statement and is not limited to the three-mode case. Any $n_{\textrm{arb}}$-mode state $\hat{\rho}_{n_{\textrm{arb}}}$ gives the same statistics, in a two-mode interferometry, as the two-mode state $\hat{\rho}_{2} = \text{Tr}_{n_{\textrm{arb}}-2}\{\hat{\rho}_{n_{\textrm{arb}}}\}$, where the trace is taken over $n_{\textrm{arb}}-2$ unused modes. 

In order to demonstrate that the measurement of populations of Zeeman components is optimal for a two-mode interferometric transformation we focus on the two $m_F = \pm 1$ states (similar calculations can be done for other combinations). The triple operators $\hat{A}_x = \hat{D}_{xy}$, $\hat{A}_y = \hat{Q}_{xy}$ and $\hat{A}_z = \hat{J}_{z}$ span the SU(2) Lie algebra with cyclic commutation relations $[\hat{A}_n,\hat{A}_k] = 2 i \epsilon_{nkl}\hat{A}_l$. The Fock state $|N_{+1}=j+m, N_0=N-2j, N_{-1}=j-m\rangle$ is a simultaneous eigenstate of $\hat{A}^2 = \sum_i \hat{A}_i^2$ with eigenvalue $4j(j+1)$, and $\hat{A}_z$ with eigenvalue $2m$. The quantum state limited to the fixed magnetization subspace can be written in the following form:
\begin{align}\label{eq:density}
\hat{\rho} = & \sum\limits_{j_1, j_2} C_{j_1, j_2}|j_1 + \bar{m}, N-2j_1, j_1 - \bar{m} \rangle \nonumber\\
& \times \langle j_2 + \bar{m}, N-2j_2, j_2 - \bar{m}|.
\end{align}  
In the two-mode interferometric transformation $\exp(-i\theta \hat{A}_y)$ the third mode $m_F = 0$ carries no information about $\theta$, thus we can work with the reduced quantum state 
\begin{align}
 \hat{\rho}_2 = \text{Tr}_0\{\hat{\rho}\} = \sum\limits_{j}C_{j,j} |j,\bar{m}\rangle \langle j,\bar{m}|,
\end{align}
where the shorthand notation $|j,\bar{m}\rangle \equiv |j+\bar{m},j-\bar{m}\rangle$ was used. The probability of measuring $N_{+1}$ atoms in the $m_F=1$ component and $N_{-1}$ atoms in the $m_F=-1$ component is equivalent to calculating $p(\{j,m\}|\theta) = \langle j,m| \hat{\rho}_2(\theta) |j,m\rangle$, where $\hat{\rho}_2(\theta) = \exp(-i\theta \hat{A}_y) \hat{\rho}_2 \exp(i\theta \hat{A}_y)$. This is a typical Mach-Zehnder interferometer, and straightforward calculations give
\begin{align}
p(\{j,m\}|\theta) = C_{j,j} [d^{(j)}_{m,\bar{m}}(2\theta)]^2 ,
\end{align}
where $d^{(j)}_{m,\bar{m}}(2\theta)=\langle j,m|e^{-i \theta \hat{A}_y} |j,\bar{m}\rangle $ is the Wigner rotation matrix. The FI is~\cite{Wasak2016}
\begin{align}
\mathcal{I}(\theta) & = 4 \sum\limits_{j,m} C_{j,j} \left[\frac{d}{d\theta} d^{j}_{m,\bar{m}} (2\theta) \right]^2 \nonumber\\
& = 4 \sum\limits_{j} C_{j,j} \langle j,m| \hat{A}_y^2 |j,m\rangle, 
\end{align}
and equals the two-mode QFI with $F_Q=4 \langle \hat{A}_y^2\rangle $ for any value of the phase $\theta$. The conclusion remains unchanged for remaining two-mode interferometers. When the interferometer involves the $m_F=0, +1$ Zeeman components, then the triple operators are 
 $\hat{A}_x=(\hat{J}_{x} + \hat{Q}_{zx})/\sqrt{2}$, 
 $\hat{A}_y=(\hat{J}_{y} + \hat{Q}_{yz})/\sqrt{2}$, 
 $\hat{A}_z=(\hat{J}_z + \sqrt{3}\hat{Y})/2$. The corresponding Fock state can be re-expressed as $|j+m,j-m,N-2j \rangle$, with $m=M+N-3j$. 
 When the interferometer involves the $m_F=0, -1$ Zeeman modes, then $\hat{A}_x=(\hat{J}_{x} - \hat{Q}_{zx})/\sqrt{2}$, $\hat{A}_y=(\hat{J}_{y} - \hat{Q}_{yz})/\sqrt{2}$, $\hat{A}_z=(\hat{J}_z - \sqrt{3}\hat{Y})/2$ and the corresponding Fock state can be re-expressed as $|N-2j,j+m,j-m \rangle$, with $m=M+N+3j$.
  
\subsubsection{Efficiency of the two-mode interferometer for an arbitrary number of input modes}\label{sub:mappingsuany}
 In the case of $n_{\rm arb }$ number of input modes, the general mixed quantum state can be written as
 \begin{equation}
  \hat{\rho} = \sum\limits_{j,j'}\sum\limits_{m,m'}\sum\limits_{\mathbf{l},\mathbf{l}'} C_{j,m,\mathbf{l}}^{j',m',\mathbf{l}'}|j+m,j-m,\mathbf{l}\rangle\langle j'+m',j'-m',\mathbf{l}'|,
 \end{equation}
 where $j\in \{0,1/2,\ldots, N/2\}$, $m\in\{-j,-j+1,\ldots,j\}$ and ${\bf l}=(l_1, \ldots, l_{n_{\rm arb }-2})$ with $\sum_{i=1}^{n_{\rm arb }-2}l_i = N-2j$. Since $n_{\rm arb }-2$ modes do not take part in the interferometry operation, we can trace them out and effectively work with
 \begin{equation}
  \hat{\rho}_{2} = \text{Tr}_{\mathbf{l}} \left[ \hat{\rho}\right] = \sum\limits_{j}\sum\limits_{m,m'}\rho_{m,m'}^{j}|j+m,j-m\rangle \langle j+m',j-m'|,
 \end{equation}
 where $\rho^{j}_{m,m'} = \sum_{\mathbf{l}}C^{j,m',\mathbf{l}}_{j,m,\mathbf{l}}$. If it happens that $\hat{\rho}$ encloses itself entirely in a subspace with fixed $m = \bar{m}$, or at least $m$ can be written as a function of $j$, i.e. $m=m(j)$, then
 \begin{equation}
  \hat{\rho}_{2} = \sum_{j} \rho^{j}_{m(j),m(j)}|j + m(j), j-m(j)\rangle\langle j+m(j), j-m(j)|,
 \end{equation}
 and it is an incoherent mixture of pure states belonging to different subspaces of $j$. Because the two-mode interferometer $\hat{A}_y$ does not connect states with different quantum number $j$, the reduced two-mode quantum Fisher information is
 \begin{equation}
  F_{Q}[\hat{\rho}_{2}] = 8\sum\limits_{j}\rho^{j}_{m(j),m(j)} [j(j+1) - m(j)^2] ,
  \end{equation}
  and based on the results from the previous Section~\ref{sub:mappingsu2} one can also show that the corresponding FI is independent of the phase $\theta$ and
  \begin{equation}
  \mathcal{I}(\theta;\hat{\rho}_{2})  = F_Q[\hat{\rho}_{2}].
 \end{equation}

\subsection{When magnetization fluctuates}

\begin{figure}[]
\begin{picture}(0,180)
\put(-130,85){\includegraphics[width=0.5\linewidth]{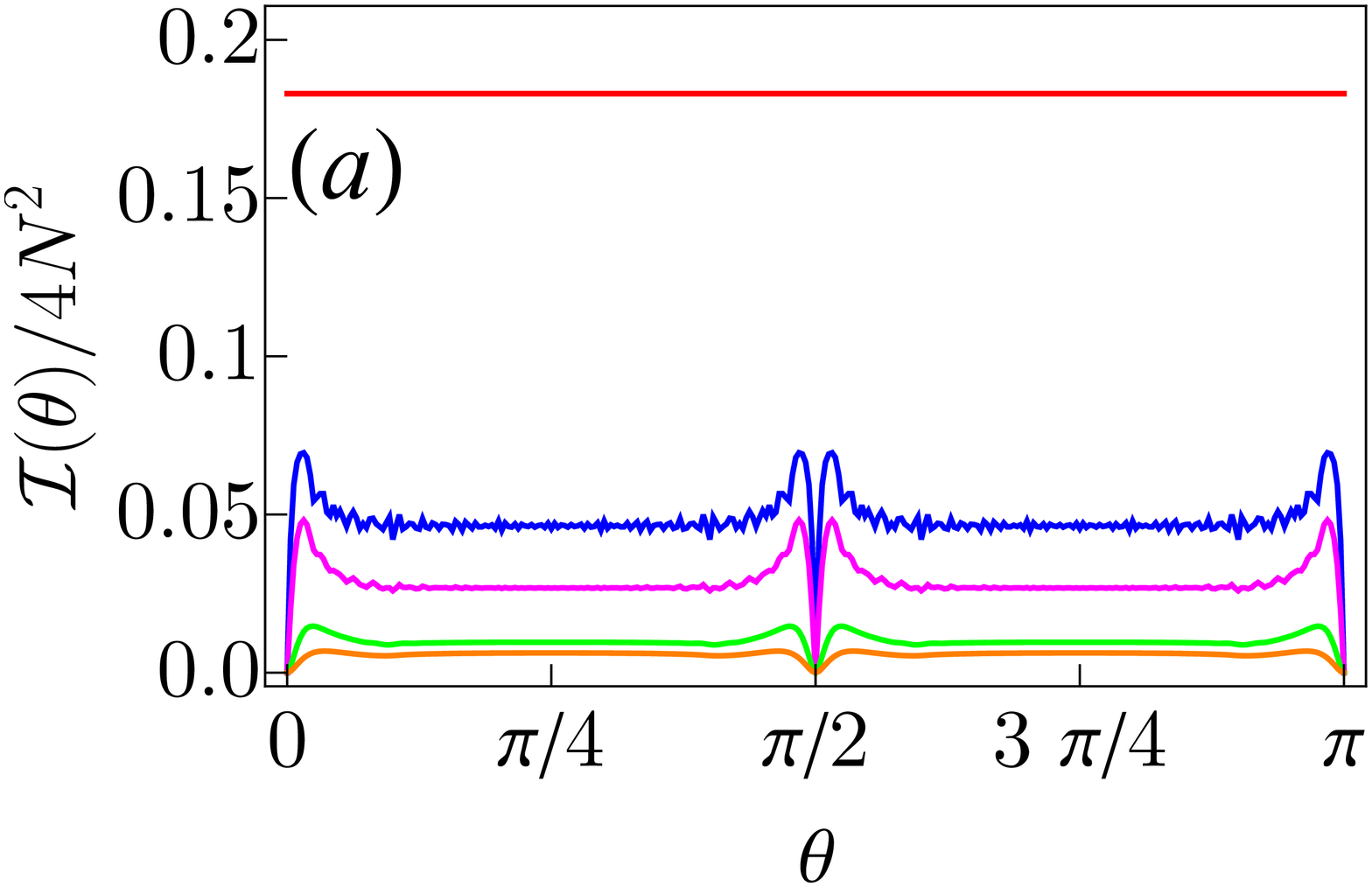}}
\put(-0,85){\includegraphics[width=0.5\linewidth]{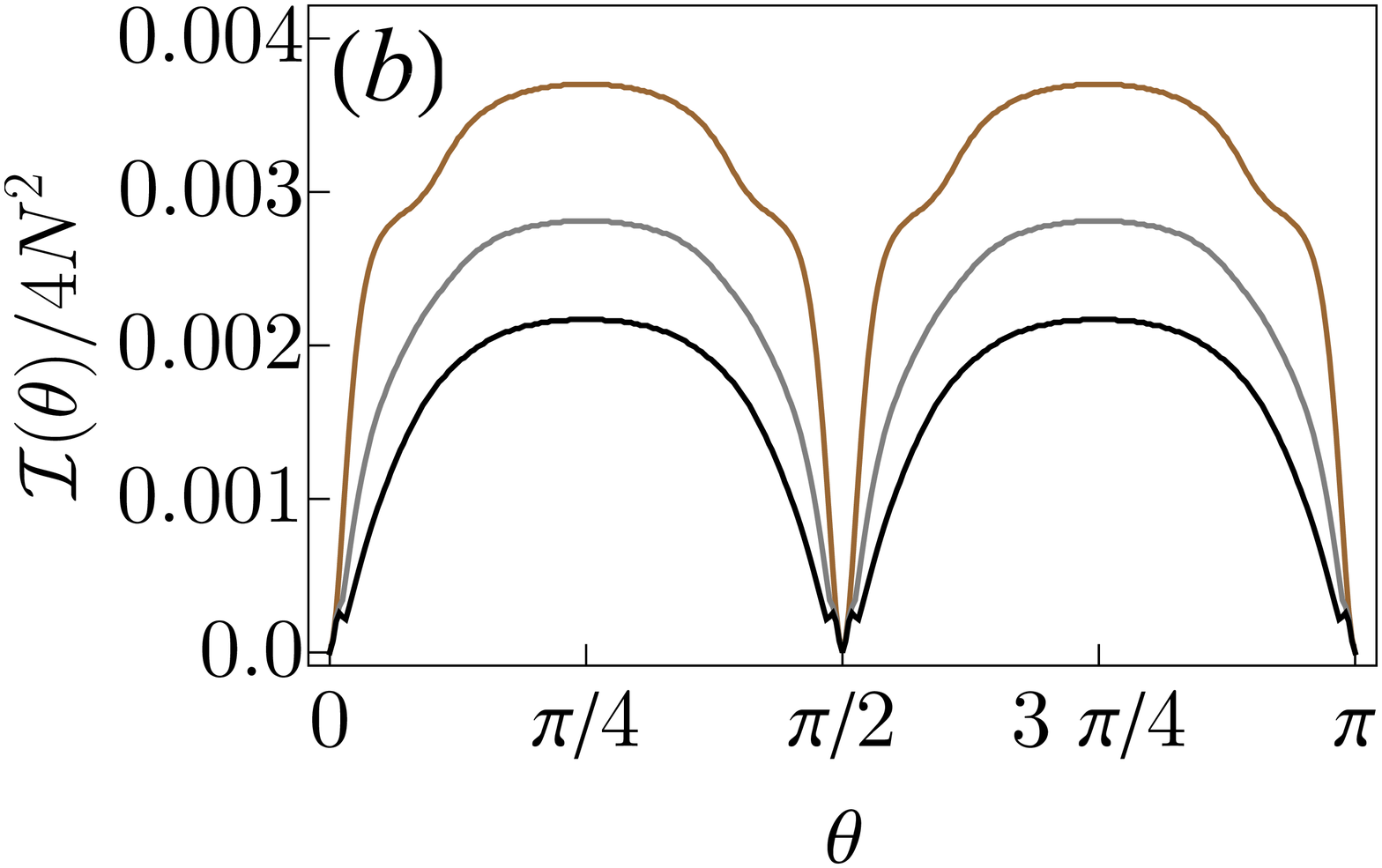}}
\put(-130,0){\includegraphics[width=0.5\linewidth]{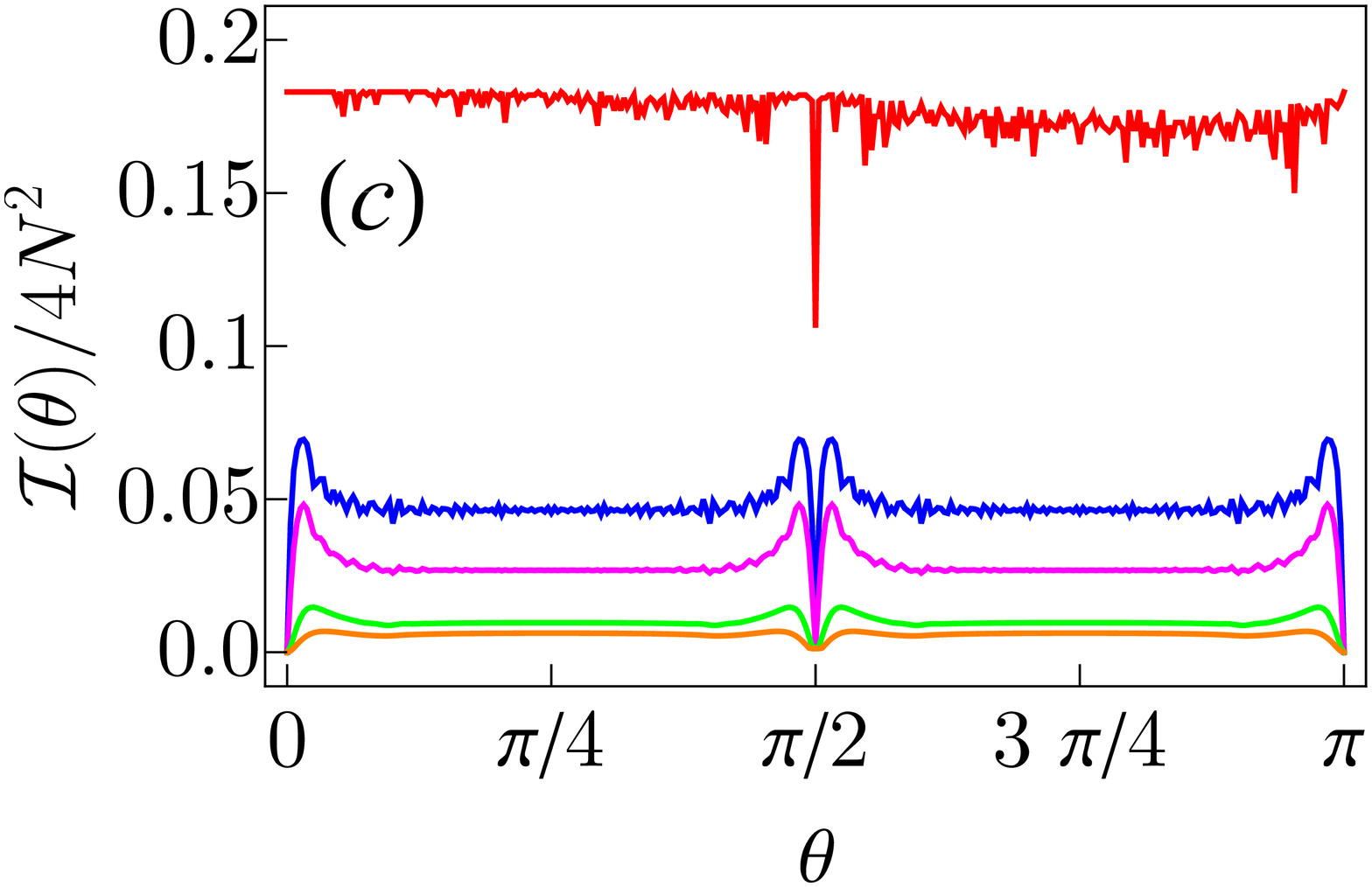}}
\put(-0,0){\includegraphics[width=0.5\linewidth]{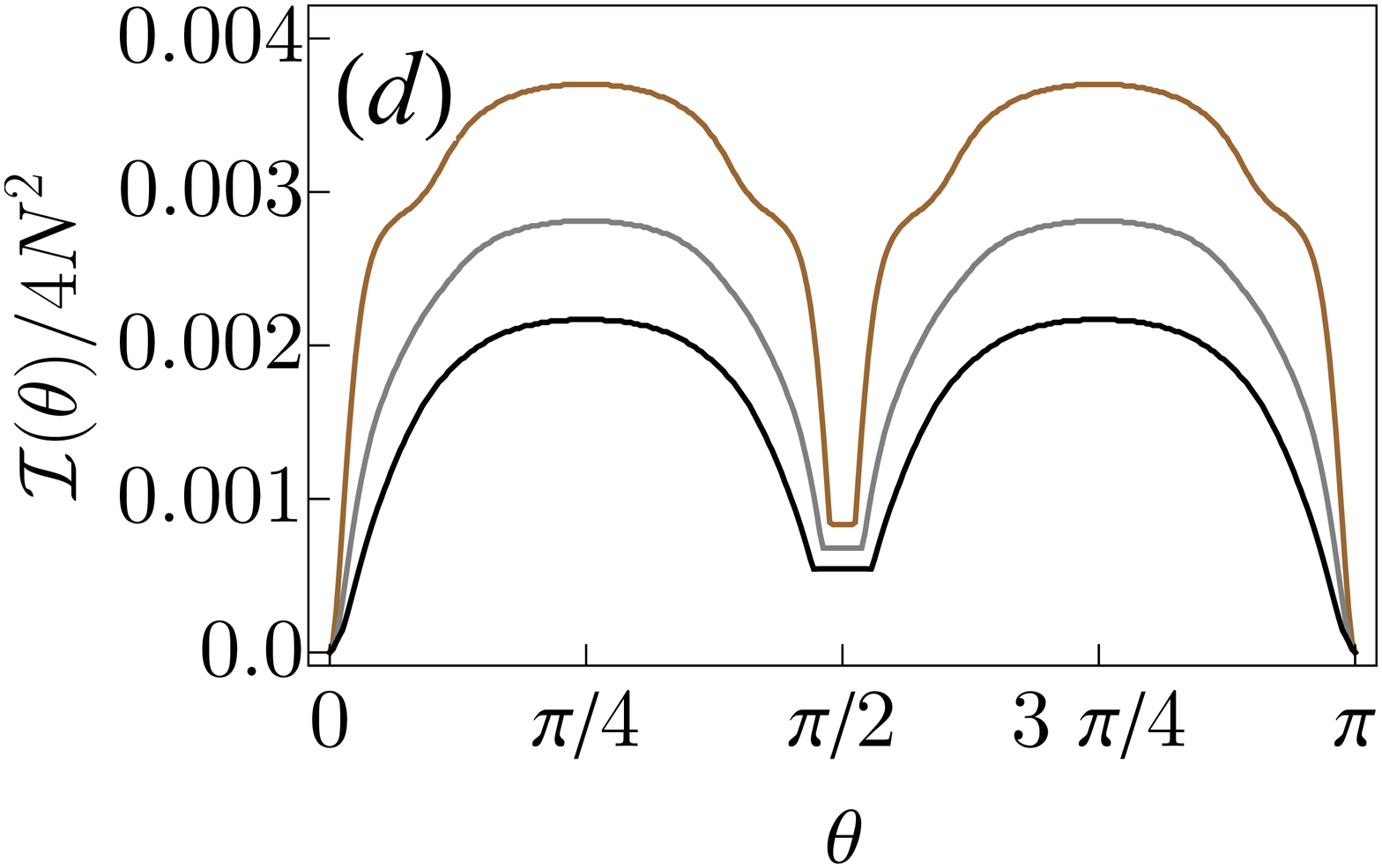}}
\end{picture}
\caption{(Color online) The FI divided by $4N^2$ from~(\ref{eq:fisher_definition}) versus $\theta$ for the two-mode with $\hat{\Lambda}_{\bf n}=\hat{K}_y$ in $(a)$-$(b)$ and the three-mode with $\hat{\Lambda}_{\bf n}=\hat{\Lambda}^{(B)}_{\bf n}$ in $(c)$-$(d)$ interferometric transformations including fluctuations of magnetization of amount: $\sigma/\sqrt{N}=0,\, \sqrt{2}/20,\, \sqrt{2}/10,\, \sqrt{2}/5,\, 3\sqrt{2}/10$ marked by colored solid lines from top to bottom in $(a)$ and $(c)$, and for $\sigma/\sqrt{N}=\sqrt{2}/2,\, 7\sqrt{2}/10,\, \sqrt{2}$ marked by colored solid lines from top to bottom in $(b)$ and $(d)$, with $N=50$, $m=0.1$, $c_2>0$, $\beta=10^{-2}$ and $q=0.5$.}
\label{fig:fig3a}
\end{figure}

\begin{figure}[]
\begin{picture}(0,200)
\put(-130,100){\includegraphics[width=0.5\linewidth]{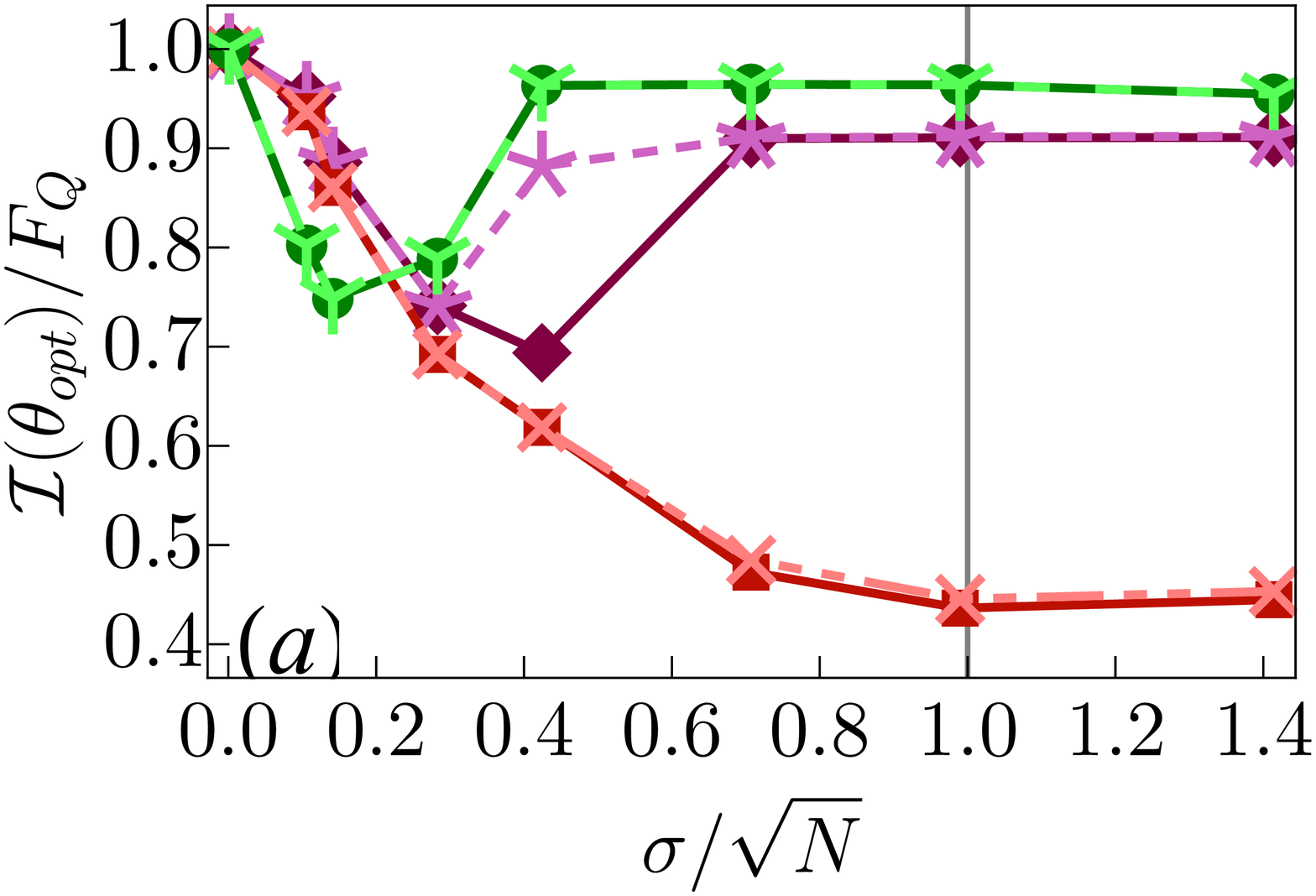}}
\put(-0,100){\includegraphics[width=0.5\linewidth]{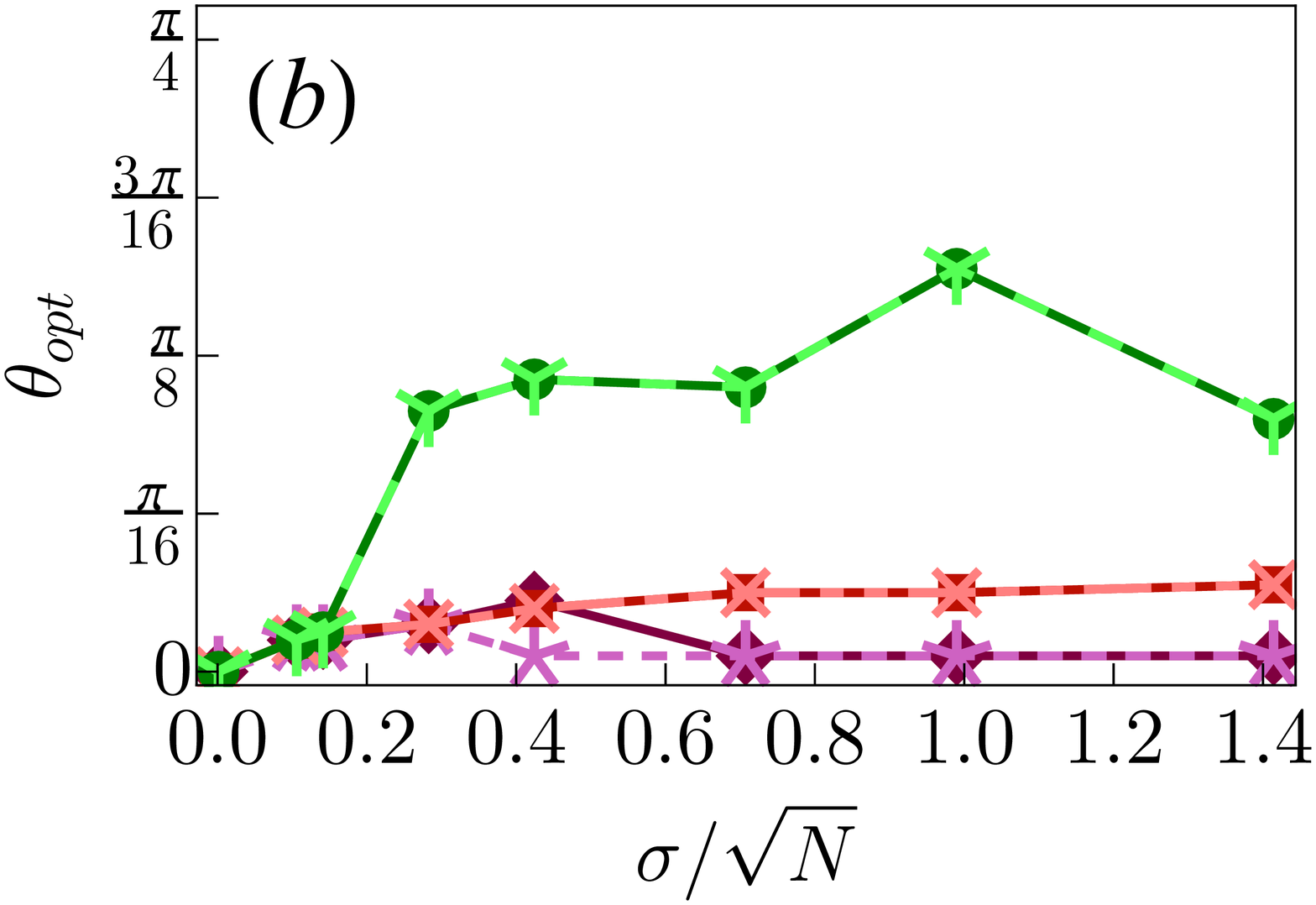}}
\put(-130,0){\includegraphics[width=0.5\linewidth]{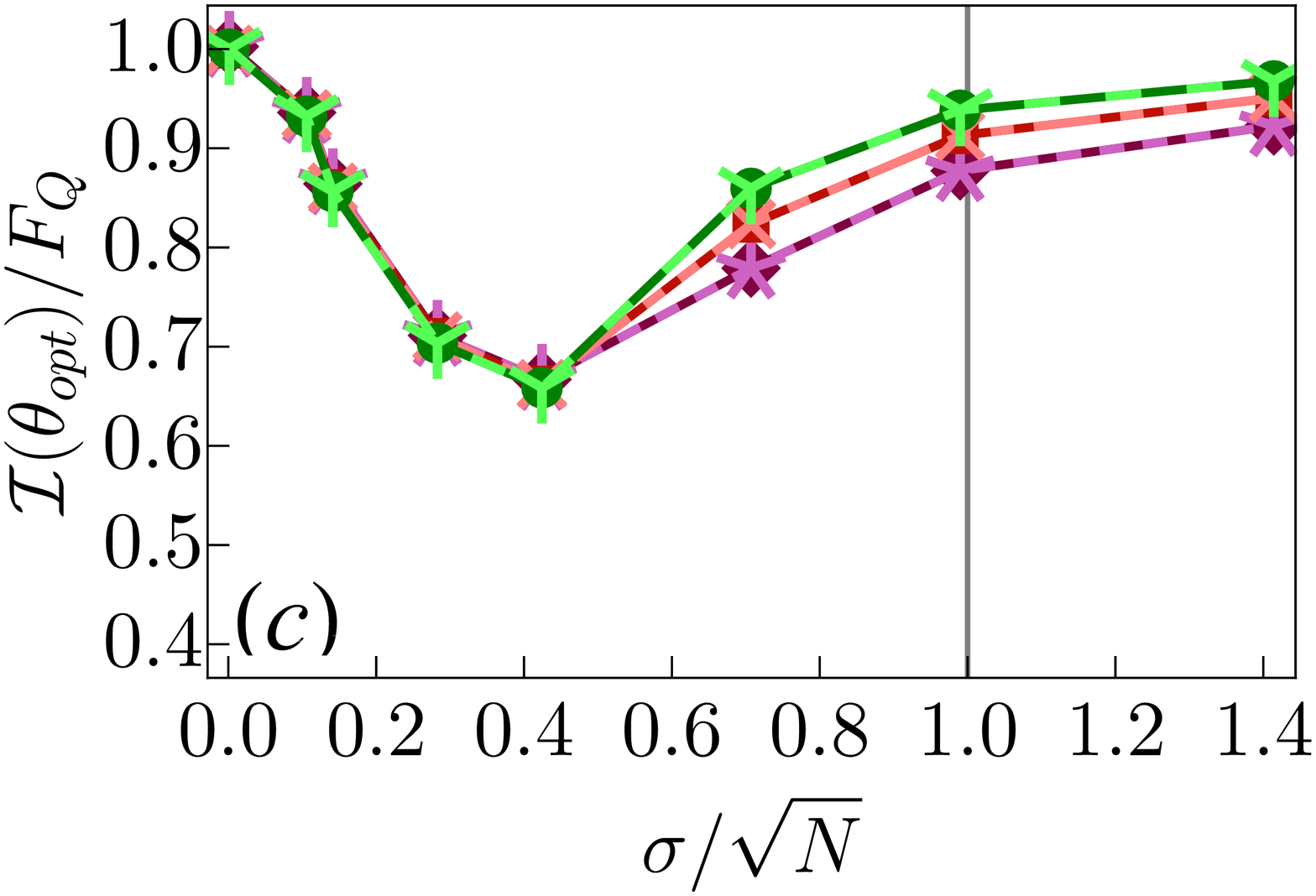}}
\put(-0,0){\includegraphics[width=0.5\linewidth]{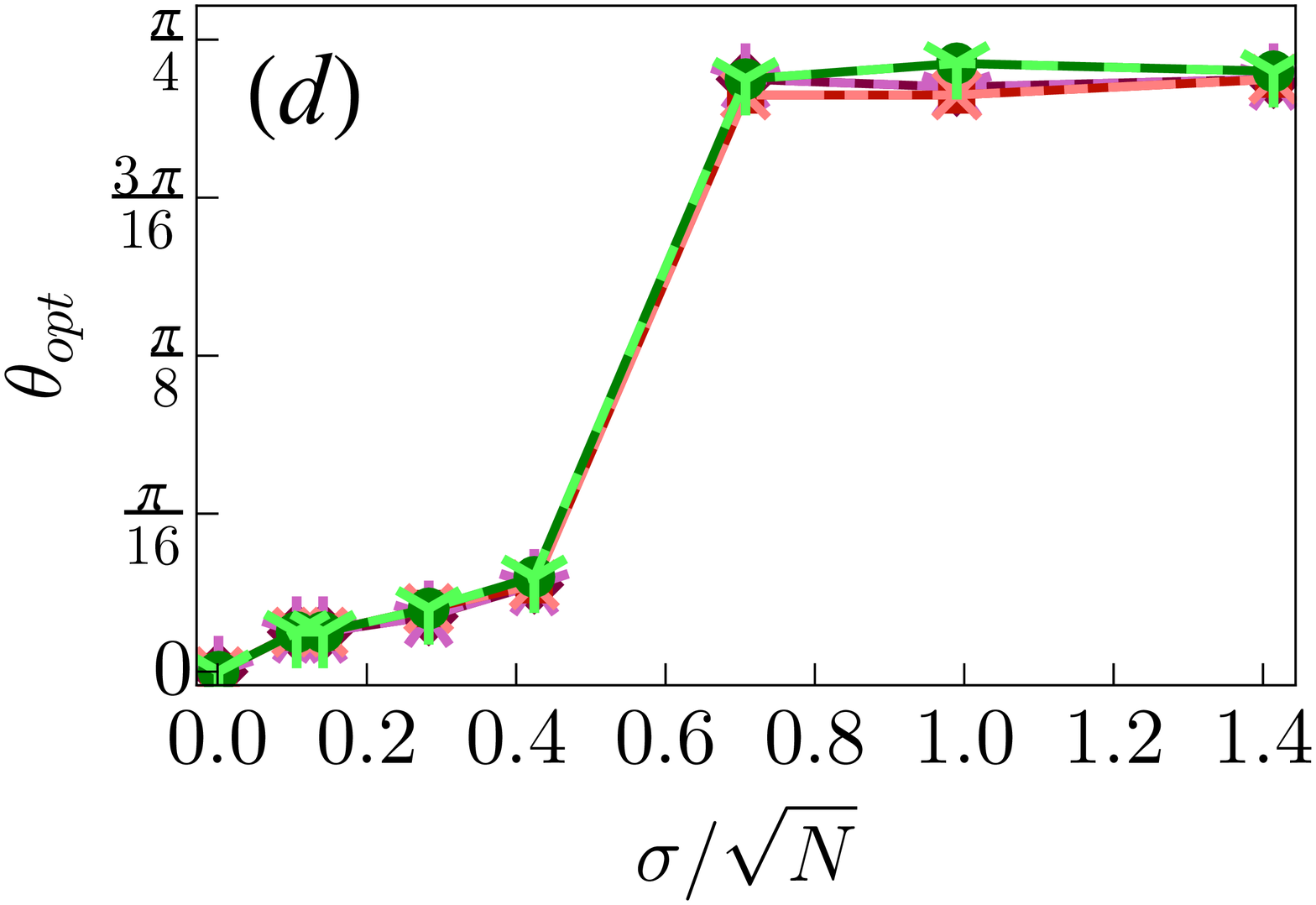}}
\end{picture}
\caption{(Color online) $(a)$ The maximal value of the relative Fisher information $\mathcal{I}(\theta)/F_Q$ versus fluctuations of magnetization $\sigma/\sqrt{N}$ for the two-mode with $\hat{\Lambda}_{\bf n}=\hat{K}_y$ (crossed points) and the three-mode with $\hat{\Lambda}_{\bf n}=\hat{\Lambda}^{(B)}_{\bf n}$ (filled points) interferometry with $q=-5$ (marked by purple stars for the two-mode and by diamonds for the three-mode cases), $q=0.5$ (marked by red crosses for the two-mode and by squares for the three-mode cases) and $q=5$ (marked by green "Y" for the two-mode and by circles for the three-mode cases), and for $\beta=10$,  $N=50$, $m=0.8, c_2>0$. 
$(b)$ $\theta_{\rm opt}$ versus $\sigma/\sqrt{N}$ for the same parameters as in $(a)$.
In $(c)$ and $(d)$ the same quantities are plotted as in $(a)$ and $(b)$, respectively, but for $\beta = 10^{-2}$. Vertical lines in $(a)$ and $(c)$ mark SQL. Colored lines linking the points are added to guide the eye.}
\label{fig:fig3b}
\end{figure}

Lets us consider the most general states for macroscopic magnetizations with non-zero fluctuations of magnetization $\Delta M=\sigma\ne 0$. The FI can be analyzed analytically for zero temperature only when the ground state in the subspace of fixed magnetization is the Fock state $|M, k_{\textrm{max}} \rangle$~\cite{PhysRevA.97.023616} for $q<0$. In this case the state (\ref{eq:rho_state}) is of the form
\begin{equation}
 \hat{\rho} = \sum\limits_{M} w_M |M,k_{\textrm{max}}\rangle\langle M,k_\textrm{max}|,
\end{equation}
A tedious evaluation of the FI at $\theta \to 0$ leads to 
\begin{align}
 \mathcal{I}(0) = & 4 \sum\limits_{M}w_M \langle M,k_\textrm{max}| \hat{\Lambda}_{\bf n}^2 | M, k_{\textrm{max}}\rangle \nonumber \\
 -& 4\sum\limits_{M',M} w_{M'}|\langle M',k_{\textrm{max}}|\hat{\Lambda}_{\bf n}|M,k_{\textrm{max}}\rangle|^2.
\end{align}
Direct calculations show that $\mathcal{I}(0) = 0$ for the optimal operator $\inter = \hat{D}_{xy}$. However, in general the variation of the FI with $\theta$ and its maximal value are unknown. As the above example indicates the optimal value of the FI may not be at $\theta = 0$ anymore. We use numerical calculations to understand the effect of magnetization fluctuations.

An example of our exact numerical results is shown in Fig.~\ref{fig:fig3a}. Indeed, the resulting FI is equal to $0$ when $\theta\to 0$, but then it increases rapidly with $\theta$, see Fig.~\ref{fig:fig3a}$(a)$-$(b)$ for the two-mode and Fig.~\ref{fig:fig3a}$(c)$-$(d)$ for the three-mode interferometers. In general, the FI depends on the phase $\theta$, and there is an optimal value of the phase, namely $\theta_{\textrm{opt}}$, for which the FI value is the largest. Thus, fluctuations of magnetization result in a shift of the optimal phase. The similar effect was also observed in the case of detection noise, as reported in~\cite{PhysRevLett.110.163604}. 

In Fig.~\ref{fig:fig3b} we gather the maximal values of the $\mathcal{I}(\theta_{\textrm{opt}})$ in $(a)$ and $(c)$, and corresponding values of $\theta_{\textrm{opt}}$ in $(b)$ and $(d)$. Unfortunately, the maximal value of the FI is not exactly equal to the QFI in some cases. It means that the measurement of populations of Zeeman components is not the most optimal one when fluctuations of magnetization start to play an important role. However, it is not the worst choice because $\mathcal{I}(\theta_{\textrm{opt}})$ still overcome the standard quantum limit as long as $\sigma<\sqrt{N}$.

\section{The special case of zero magnetization}\label{sec:zeromagnetization}
The authors mentioned in~\cite{PhysRevA.97.023616} that the optimal value of the QFI is determined by the maximum among the three values: $\lambda_A$, $\lambda_B$ and also $\lambda_C=\Gamma_{77}$ in general. The eigenvalue $\lambda_C$ corresponds to the optimal interferometric transformation $\hat{\Lambda}_{\bf n}^{(C)}=\hat{Y}$. In the case of macroscopic magnetization the operators $\inter^{(A,B)}$ were sufficient to determine the optimal value of the QFI for all ranges of temperature and magnetic field strength. This is no longer true when $M$ is close to zero in the high temperature limit, where fluctuations of $\inter^{(C)}$ become dominant. In this special case, the QFI of the ground state ($\beta \to \infty$) is determined by $\lambda_A$ in the Twin-Fock phase (also called antiferromagnetic) for $q < q_c$ and $\lambda_B$ in the polar and broken-axisymmetry phase for $q>q_c$, where $q_c$ is the critical point which at $M= 0$ is $q_c=0$ for $c_2>0$, and $q_c=-2$ for $c_2<0$. The situation changes in the high temperature limit ($\beta \to 0$) when the QFI is determined by the value of $\lambda_C$ only.

\begin{figure}[]
\begin{picture}(0,80)
\put(-130,0){\includegraphics[width=0.5\linewidth]{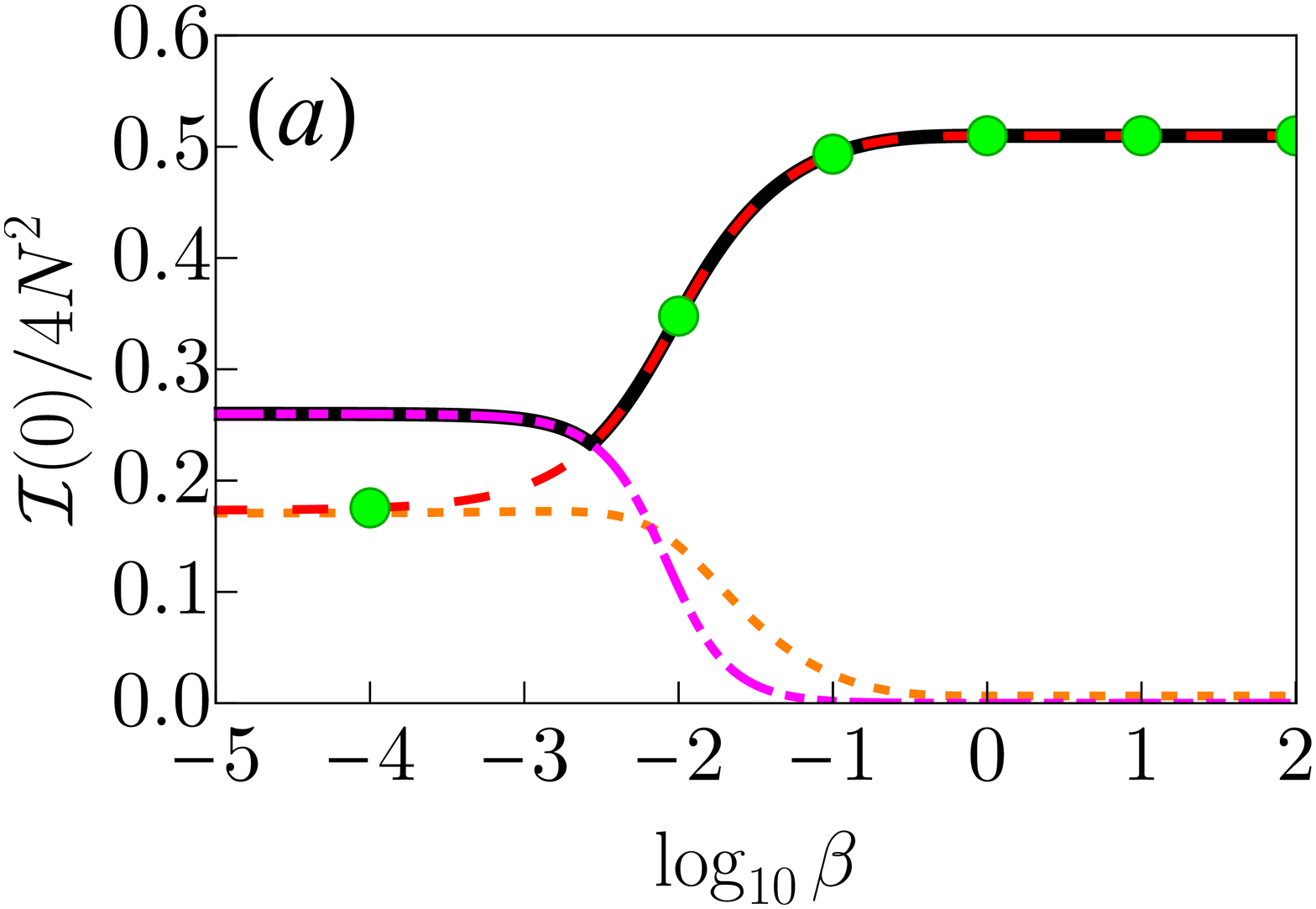}}
\put(-0,0){\includegraphics[width=0.5\linewidth]{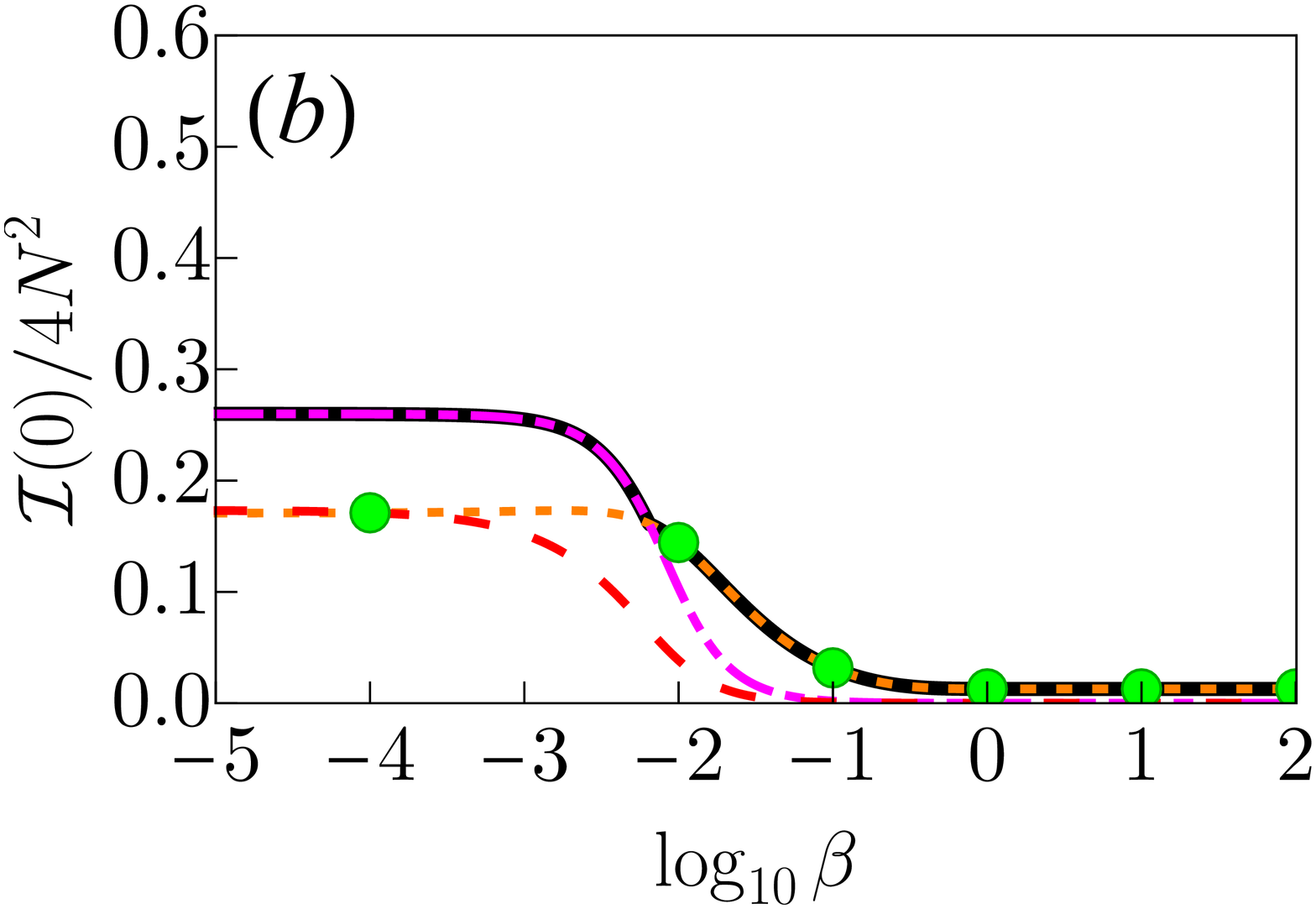}}
\end{picture}
\caption{(Color online) The FI at $\theta=0$ (green points) versus  $\log_{10}\beta$ for $q=-5$ in $(a)$ and for  $q=5$ in $(b)$ compared to the QFI (black solid line), with $N=100$, $c_2<0$, $\sigma=0$, $M=0$. The variances of $\hat{D}_{xy}$ (red dashed line), $\hat{J}_x$ (orange dotted line) and $\hat{Y}$ (purple dot dashed line) are also shown. All quantities are divided by $4N^2$.}
\label{fig:fig0}
\end{figure}

In Fig.~\ref{fig:fig0} we show an example of exact numerical results for the FI (points) based on the measurement of populations of Zeeman components compared to the QFI (black solid lines) in the case of zero fluctuations of magnetization ($\sigma\to 0$). 
As long as the temperature is low ($\beta\to \infty$), the measurement of Zeeman components populations is optimal, i.e. maximizes the FI. In the high temperature limit ($\beta \to 0$) the value of $\lambda_C$ is the highest and determines the value of the QFI. Unfortunately, operator $\hat{Y}$ is diagonal in the Fock state basis and particle number measurement carries no information about $\theta$ in the case of the interferometer $\inter^{(C)}$ yielding $\mathcal{I}(\theta) = 0$. Hence, other rotations with ($\hat{\Lambda}_{\bf n}^{(A)}$ or $\hat{\Lambda}_{\bf n}^{(B)}$) are a better choice because their variance scales as $N^2$. 

Interesting situation takes place in the polar phase for $q > 2$. In the ground state all atoms occupy the $m_F = 0$ Zeeman state, forming a coherent state $|0,N,0\rangle$ with the $F_Q = 4N$. When the temperature grows, higher energy entangled states are being populated increasing the QFI value, as illustrated in Fig.~\ref{fig:fig0}$(b)$. Finally, in the high-temperature limit ($\beta \to 0$) the QFI gains the Heisenberg scaling. 
The careful reader can notice the same effect in Fig.~\ref{fig:fig2}$(d)$ for macroscopic magnetizations, where the temperature slightly increases the value of the QFI. A similar effect was recently reported in~\cite{PhysRevA.97.032339}.

\section{Precision from the measurement of signal}\label{sec:moments}

The method of moments~\cite{Pezze2016-dx,Pezze2014,1751-8121-47-42-424006} is an alternative estimation strategy, which does not require the knowledge of the probability distribution $p(x|\theta)$. The strategy can be of less experimental effort, provided the signal is proper and possible to measure. The purpose of this section is to identify an appropriate observable leading to the signal possible to measure for the states (\ref{eq:rho_state}) with fixed magnetization.

The observable $\hat{\mathcal{P}}$ with $\theta$ dependent expectation value $\langle \hat{\mathcal{P}} \rangle_{\theta}$ carries information about unknown value of $\theta$, and thus can be exploited in the estimation procedure. In the limit of a large number of measurements $\nu \gg 1$ the estimation precision $\Delta\theta_{\textrm{mom}}$ is given by the error-propagation formula~\cite{1751-8121-47-42-424006,Pezze2016-dx,Pezze2014}: $\Delta \theta_{\textrm{mom}} = (\Delta \hat{\mathcal{P}})_{\theta}/|\sqrt{\nu} \partial_\theta \langle \hat{\mathcal{P}} \rangle_{\theta}|$, where $(\Delta \hat{\mathcal{P}})_{\theta} = \sqrt{\langle \hat{\mathcal{P}}^2 \rangle_{\theta} - \langle \hat{\mathcal{P}}\rangle_{\theta}^2}$. The uncertainty of $\theta$ is bounded from below by the QFI 
\begin{equation}\label{eq:unthetaFq}
 (\sqrt{\nu}\Delta \theta_{\textrm{mom}})^{-2} \leq F_Q.
\end{equation}

In the two-mode interferometry introduced in Section~\ref{sub:mappingsu2}, the input states are effectively incoherent mixtures of Dicke states, i.e. $\hat{\rho} = \sum_{j} C_{j,j}|j,m\rangle \langle j,m|$, where $|j,m\rangle  \equiv |j+m,j-m\rangle$ is a short-hand notation for the two-mode Fock state with $j+m$ atoms in a state $a$ and $j-m$ in a state $b$. From now on, we drop any reference to specific Zeeman states and work with general creation operators $\hat{a}^{\dagger}$ and $\hat{b}^{\dagger}$ and spin operators 
\begin{subequations}
 \begin{align}
  \hat{S}_x = & \frac{1}{2}(\hat{a}^{\dagger}\hat{b} + \hat{b}^{\dagger}\hat{a}), \\
  \hat{S}_y = & \frac{1}{2i}(\hat{a}^{\dagger}\hat{b} - \hat{b}^{\dagger}\hat{a}), \\
  \hat{S}_z = & \frac{1}{2}(\hat{a}^{\dagger}\hat{a} - \hat{b}^{\dagger}\hat{b}), 
 \end{align}
\end{subequations}
which satisfy cyclic commutation relations $[\hat{S}_n, \hat{S}_k] = i \epsilon_{nkl}\hat{S}_l$. It was noted in~\cite{Iagoba2015} that the measurement of the $\hat{S}_z^2$ operator saturates the inequality~\eqref{eq:unthetaFq} for a specific value of $\theta$. As we will demonstrate, this result holds only for Dicke states with $m=0$. Otherwise, the precision $\Delta\theta_{\textrm{mom}}$ diverges rapidly with $m$ from its optimal value given by the square root inverse of the QFI. 
We also show that the measurement of the parity operator $\hat{\Pi}_b$ is not optimal for general incoherent mixture, however it saturates the inequality~\eqref{eq:unthetaFq} in the case if a pure Dicke state.

\subsection{Measurement of the $\hat{S}_z^2$ operator}
\begin{figure}
 \centering
 \begin{adjustbox}{width=0.95\linewidth,height=0.7\linewidth}
 \begin{tikzpicture}
  \pgfmathsetmacro{\j}{10}
  \begin{axis}[
    axis x line=middle,    
    axis y line=middle,    
    axis line style={-latex}, 
    extra x ticks=0,
    xlabel={$m$},          
    ylabel={$(\sqrt{\nu}\Delta\theta_{\rm min})^{-2}/4j^2$},          
    xmin=-12, 
    xmax=12,
    ymin=0.0, 
    ymax=0.65,
  	samples=500,
    domain=-\j:\j,
    no markers]
   \addplot[dashed, blue] plot ({\x},{2*(\j*(\j+1)-\x*\x)/(4*\j*\j)});
   \addplot[solid, red] plot ({\x},{2*pow((\j*(\j+1)-3*\x*\x),2)/(\j*(\j+1)*(1+4*\x*\x)-\x*\x*(5+4*\x*\x))*1/(4*\j*\j)});
  \end{axis}
 \end{tikzpicture}
 \end{adjustbox}
 \caption{(Color online) The precision from the error-propagation formula (\ref{eq:Sz2}) (red solid) for a single Dicke state $\hat{\rho} = |j,m\rangle\langle j,m|$ versus the fractional magnetization $m$ compared to the lower bound set by the QFI (blue dashed). In the figure, $j = 10$.}
 \label{fig:Dicke_comparison}
\end{figure}
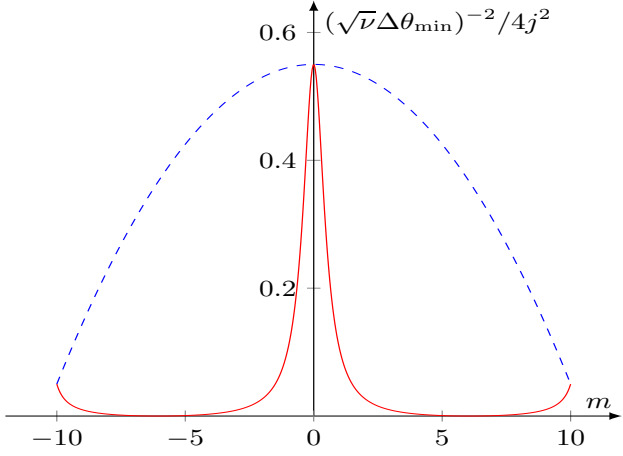

In the Mach-Zender interferometer information about $\theta$ is imprinted on the initial state $\hat{\rho}$ through a unitary transformation $\hat{\rho}_{\theta} = \hat{U}_{\theta} \hat{\rho} \hat{U}_{\theta}^{\dagger}$, where $\hat{U}_{\theta} = \exp(-i\theta \hat{S}_y)$. The expectation value of $\hat{S}_z^2$ equals
\begin{align}
\langle \hat{S}_z^2 \rangle_{\theta}
=& \text{Tr}\left\{\hat{S}_z^2 e^{-i\theta \hat{S}_y} \hat{\rho} e^{i\theta \hat{S}_y} \right\} \nonumber \\
=& \cos^2\theta \langle \hat{S}_z^2 \rangle + \sin^2\theta \langle \hat{S}_x^2 \rangle - \cos\theta \sin\theta \langle \{\hat{S}_z, \hat{S}_x\} \rangle,
\label{eq:unc}
\end{align}
where we used the formula $\hat{U}_{\theta}^{\dagger} \hat{S}_z \hat{U}_{\theta} = \cos\theta \hat{S}_z - \sin\theta \hat{S}_x$.
In the method of moments uncertainty of $\theta$ follows from uncertainty of $\langle \hat{S}_z^2\rangle_{\theta}$~\cite{Pezze2016-dx} and takes the form 
\begin{equation}
 \Delta \theta_{\textrm{mom}} = \frac{(\Delta \hat{S}_z^2)_\theta}{\sqrt{\nu} \left| \frac{d}{d\theta}\langle \hat{S}_z^2\rangle_{\theta}\right|}, \ \ \ \ {\rm for }\,\, \nu \gg 1.
\end{equation}
As explained in~\cite{Iagoba2015}, and also in Appendix~\ref{ap:calculsz2}, the uncertainty $\Delta \theta_{\textrm{mom}}$ for a special class of quantum states $\hat{\rho}$ becomes  
\begin{align}
& \nu \Delta\theta_{\textrm{mom}}^2  =  \nonumber \\
& \frac{(\Delta \hat{S}_x^2)^2 f(\theta) + 4\langle\hat{S}_x^2\rangle - 3\langle \hat{S}_y^2\rangle - 2\langle \hat{S}_z^2\rangle(1 + \langle \hat{S}_x^2\rangle) + 6\langle\hat{S}_z\hat{S}_x^2\hat{S}_z\rangle}{4(\langle\hat{S}_x^2\rangle - \langle\hat{S}_z^2\rangle)^2},
\end{align}
where
\begin{equation}
f(\theta)  = \left[\frac{(\Delta \hat{S}_z^2)^2}{(\Delta \hat{S}_x^2)^2}\frac{1}{\tan^2\theta} + \tan^2\theta \right].
\end{equation}
The phase $\theta_{\textrm{min}}$ which minimizes the uncertainty (\ref{eq:unc}) is
\begin{equation}
 \tan^2\theta_{\textrm{min}} = \frac{\Delta \hat{S}_z^2}{\Delta \hat{S}_x^2},
\end{equation}
thus the estimation procedure performed around the value $\theta_{\textrm{min}}$ gives the optimal precision $\Delta \theta_{\textrm{min}}$ with
\begin{align}
& \nu \Delta\theta_{\textrm{min}}^2 = \nonumber \\
& \frac{2\Delta\hat{S}_z^2 \Delta \hat{S}_x^2+ 4\langle\hat{S}_x^2\rangle - 3\langle \hat{S}_y^2\rangle - 2\langle \hat{S}_z^2\rangle(1 + \langle \hat{S}_x^2\rangle) + 6\langle\hat{S}_z\hat{S}_x^2\hat{S}_z\rangle}{4(\langle\hat{S}_x^2\rangle - \langle\hat{S}_z^2\rangle)^2}. 
\end{align}
For an incoherent mixture of Dicke states $\hat{\rho} = \sum_j C_{j,j}|j,m\rangle \langle j,m|$, introduced in Section~\ref{sub:mappingsu2}, we have:
\begin{align}
 \langle \hat{S}_x^2\rangle & = \langle \hat{S}_y^2\rangle = \frac{1}{2}\sum\limits_{j}C_{j,j}  [j(j+1)] - \frac{1}{2}m^2,\\
 \langle \hat{S}_z^2 \rangle & = m^2, \\
  \Delta \hat{S}_z^2 & =0, \\
 \langle \hat{S}_z \hat{S}_x^2 \hat{S}_z \rangle & = m^2 \langle \hat{S}_x^2 \rangle,
\end{align}
and the inverse of precision squared is equal to
\begin{equation}\label{eq:Sz2}
 (\sqrt{\nu}\Delta \theta_{\textrm{min}})^{-2} =  \frac{2\left[\sum_j C_{j,j} j(j+1)-3m^2\right]^2}{\left[\sum_j C_{j,j}j(j+1)-m^2\right](1+4m^2)-4m^2}. 
\end{equation}
The above expression agrees with the QFI value $F_{Q}[\hat{\rho}] = 2 \sum_j C_{j,j}j(j+1) - 2m^2$ only if $m$ is zero. Otherwise, it diverges rapidly with $m$, as can be seen from the Taylor expansion around small $m$ (see Fig.~\ref{fig:Dicke_comparison} for a special case $\hat{\rho} = |j,m\rangle \langle j,m|$).
Therefore, the measurement of $\hat{S}_z^2$ is not the optimal one for the states with macroscopic magnetization and $m\ne 0$.

\subsection{Mach-Zehnder interferometry with the parity measurement}
The parity operator is diagonal in the particle number basis~\cite{PhysRevA.54.R4649}
\begin{equation}
 \hat{\Pi} = (-1)^{\hat{N}_b} = \sum\limits_{j,m} (-1)^{j-m}|j,m\rangle\langle j,m|,
\end{equation}
and has only two eigenvalues $\pm 1$ depending on the particle number parity in the $b$ component. This feature makes it very sensitive to e.g. detection noise~\cite{PhysRevLett.110.163604}. The parity operator can be measured by counting the number of particles in one of the components and assigning to this result the eigenvalue $\pm 1$, depending on the parity. This requires single particle resolution~\cite{Gerry2010,Ott2016}.

The expectation value $\langle \hat{\Pi} \rangle_{\theta}$ of the parity operator calculated with the initial mixture of Dicke states $\hat{\rho} = \sum_j C_{j,j}|j,m\rangle\langle j,m|$ takes the form
\begin{align}\label{eq:parity}
 \langle \hat{\Pi}\rangle_{\theta} &= \sum\limits_{j} C_{j,j} \langle j,m| e^{i\theta\hat{S}_y} \cdot \hat{\Pi} \cdot e^{-i\theta\hat{S}_y} |j,m\rangle \nonumber \\
 {}&= \sum\limits_{j} C_{j,j} \sum\limits_{m'=-j}^{j}(-1)^{j-m'} \left|\langle j,m| e^{i\theta\hat{S}_y}|j,m'\rangle\right|^2 \nonumber \\
 {}&= \sum\limits_{j}C_{j,j}\sum\limits_{m'=-j}^{j}(-1)^{j-m'}\left[d^{j}_{m,m'}(\theta) \right]^2.
\end{align}
Since $\hat{\Pi}^2 = \mathds{1}$, the error-propagation formula reduces to
\begin{equation}\label{eq:parity_error_propagation}
 (\Delta\theta_{\textrm{mom}}\sqrt{\nu})^{-2} = \frac{\left| \frac{d \langle\hat{\Pi}\rangle_{\theta}}{d\theta} \right|^2}{1 - \langle\hat{\Pi}\rangle_{\theta}^2} \leqslant F_Q[\hat{\rho}]
\end{equation}
In general, it is not possible to saturate the inequality in Eq.~\eqref{eq:parity_error_propagation} when at least two coefficients $C_{j,j}$ are non-zero. However, in a special case $\hat{\rho} = |j,m\rangle\langle j,m|$ of a pure Dicke state the $0/0$ expression appears in~\eqref{eq:parity_error_propagation} at $\theta = 0$. The Taylor expansion of the expectation value~\eqref{eq:parity} to the 4-th order around small $\theta$ gives
\begin{align}
 \langle\hat{\Pi}\rangle_{\theta} = & (-1)^{j-m} \left[1 - \theta^2 \cdot P_2(j,m) \right.\nonumber\\
 &\left.+ \theta^4 \cdot P_4(j,m) \right],
\end{align}
where 
\begin{align}
 P_2(j,m) & = \langle j,m|\hat{S}_y^2|j,m\rangle + \sum\limits_{m'} (-1)^{m-m'}\langle j,m|\hat{S}_y|j,m'\rangle^2 \nonumber \\
 &= 2[j(j+1) - m^2] = \frac{F_Q}{2}, \\
 P_4(j,m) & = \frac{1}{12}\langle j,m|\hat{S}_y^4|j,m\rangle \nonumber \\
 & + \frac{1}{3}\sum\limits_{m'}(-1)^{m-m'}\langle j,m|\hat{S}_y|j,m'\rangle \langle j,m|\hat{S}_y^3|j,m'\rangle \nonumber \\
 &+ \frac{1}{4} \sum\limits_{m'}(-1)^{m-m'}\langle j,m|\hat{S}_y^2|j,m'\rangle^2.
\end{align}
From the error-propagation formula~\eqref{eq:parity_error_propagation} we get 
\begin{align}
 (\Delta\theta_{\textrm{mom}} \sqrt{\nu})^{-2} &= F_Q + \theta^2\left[\frac{F_Q^2}{4} - 6P_4(j,m)\right] +\mathcal{O}(\theta^3).
\end{align}
The measurement of parity around $\theta = 0$ is optimal for any Dicke state $|j,m\rangle$, irrespective of $m$. In addition, for the Twin-Fock state, with $m=0$, estimation around $\theta=\pi/2$ is also optimal. 

\section{Conclusions}
Usefulness of spinor Bose-Einstein condensates for atomic interferometry is investigated experimentally nowadays~\cite{{NaturePhys8},{PhysRevLett.115.163002},{10.2307/41351684},{TFScience2017},ChapmanPNAS,PhysRevLett.117.143004, Peise2015}, including observation
of the twin Fock state~\cite{{10.2307/41351684},{TFScience2017}}. However, main efforts concentrate around the special case of zero magnetization. Our results show that focusing on a specific value of magnetization is very limiting. 

In this paper we focused on the calculation of the Fisher information for the spin-1 Bose-Einstein condensates with thermally populated internal degrees of freedom, and show that the measurement of number of atom population in particular Zeeman components maximizes its value. We introduced the concept of the effective two-mode interferometry in the three-mode system. When the information about the parameter $\theta$ is contained entirely within two modes, then quantum metrological properties of the three-mode state are the same as of the two-mode state obtained as a trace over an unexploited mode. The same mapping can be performed for an arbitrary number $n_{\rm arb}$ of modes and the corresponding SU($n_{\rm arb}$) interferometer. In other words, the two-mode interferometry can be implemented effectively in the systems consist of atoms with higher spin provided that the interferometric transformation involves only two modes. The two-mode Fisher information is independent of $\theta$ and has Heisenberg scaling as long as the magnetization variance is smaller than 1 even for non-zero temperatures. Moreover, the temperature can be a source of increasing the Fisher information as we illustrated in Section \ref{sec:zeromagnetization} for the coherent state arising for zero magnetization in the high magnetic field limit. Fluctuations of magnetization make the two-mode Fisher information $\theta$ dependent, and hence introduce its optimal value for which the precision in the estimation is the best.  

Our results revealed a great potential of spinor condensates for quantum interferometry not only for zero magnetization but also for a macroscopic one. However, using them in practice can be of the same efficiency as coherence states because decoherence effects or detection noise prevent take advantage of their properties \cite{PhysRevX.7.041009}. It is the fact for other entangled states as well. However, in the light of recent theoretical and experimental results \cite{PhysRevLett.116.053601, PhysRevLett.119.193601, PhysRevLett.117.013001} it is interesting to develop an alternative interferometric protocol that could diminish or even reduce such destructive effects. This provides an interesting directions for a further work.

\acknowledgments
We acknowledge discussion with K. Paw\l{}owski. This work was supported by the Polish National Science Center through Grants no. 2012/07/E/ST2/01389 and DEC-2015/18/E/ST2/00760.

\appendix

\section{Equivalent expressions for the Fisher information}\label{ap:FI}

According to the general definition (\ref{eq:fisher_definition}), the Fisher information can be calculated also from the following expressions:
\begin{align}
\mathcal{I}(\theta) & = 
\sum\limits_{M,k} \frac{1}{p(\{M,k\}|\theta)}\left(\frac{\partial p(\{M,k\}|\theta)}{\partial\theta}\right)^2 \nonumber \\
{}&=\sum\limits_{M,k} \left(\frac{\partial \log p(\{M,k\}|\theta)}{\partial\theta} \right)^2 p(\{M,k\}|\theta)  \\
{}&=4 \sum\limits_{M,k} \left(\frac{\partial \sqrt{p(\{M,k\}|\theta)}}{\partial\theta}\right)^2 \label{eq:fishersqrt}\\
& = -\sum\limits_{M,k}p(\{M,k\}|\theta) \frac{\partial^2}{\partial\theta^2}\log p(\{M,k\}|\theta).
\end{align}

\section{Numerical procedure for the Fisher information calculations}\label{ap:numerics}

The Fisher information was calculated numerically based on Eq.~(\ref{eq:fishersqrt}) by rotation of the density matrix written in the Fock-state representation. In this way, one considers only the representation of operators and states in a more familiar vector space. 
Operations such as dot product, addition and multiplication transfer into the vector space, observables are represented by square hermitian matrices, and ket states as column vectors.
Eigenvectors and eigenvalues of the Hamiltonian written in the Fock state basis were used to form the density matrix (\ref{eq:density}). They were calculated with algorithms built in the MATLAB enviroment, as well as rotation of the density matrix. 

\section{Expressions for the error propagation formula in the case of $\langle \hat{S}_z^2 \rangle$ measurement}\label{ap:calculsz2}

After some algebra one gets
\begin{equation}
 \frac{d}{d\theta}\langle\hat{S}_z^2 \rangle_{\theta} = \sin(2\theta) \left[ \langle\hat{S}_x^2\rangle - \langle\hat{S}_z^2 \rangle\right] -\cos(2\theta) \langle\{\hat{S}_z, \hat{S}_x \} \rangle,
\end{equation}
 and
\begin{align} 
 &(\Delta \hat{S}_z^2)_{\theta}^2  = (\Delta \hat{S}_z^2)^2 \cos^4\theta + (\Delta \hat{S}_x^2)^2 \sin^4\theta \nonumber \\
 & + \cos\theta\sin^3\theta \left[2\langle\hat{S}_x^2\rangle\langle\{ \hat{S}_z,\hat{S}_x\}\rangle - \langle\{\hat{S}_x^2,\{\hat{S}_x,\hat{S}_z \} \}\rangle \right] \nonumber\\
 & + \cos^3\theta\sin\theta \left[ 2\langle\hat{S}_z^2\rangle\langle\{ \hat{S}_z,\hat{S}_x\}\rangle - \langle\{\hat{S}_z^2,\{\hat{S}_x,\hat{S}_z \} \}\rangle \right] \nonumber\\
 & +\cos^2\theta\sin^2\theta \left[\langle \{ \hat{S}_z, \hat{S}_x\}^2\rangle - \langle \{ \hat{S}_z, \hat{S}_x\} \rangle^2 \right. \nonumber \\
 &+\left. \langle\{\hat{S}_x^2,\hat{S}_z^2\}\rangle-2\langle\hat{S}_z^2\rangle\langle\hat{S}_x^2\rangle \right],
\end{align}
where $\{ \cdot, \cdot \}$ denotes anti-commutator.
In the case of quantum states within a fixed magnetization subspace the above formulas simplify due to
\begin{align}
 \langle \{\hat{S}_z, \hat{S}_x \} \rangle & = 0, \\
 \langle \{ \hat{S}_x^2, \{\hat{S}_z, \hat{S}_x \}\} \rangle & = 0, \\
 \langle \{ \hat{S}_z^2, \{\hat{S}_z, \hat{S}_x \}\} \rangle & = 0,
\end{align}
and one recovers the result of~\cite{Iagoba2015}.

\section{Generalized Wigner rotation matrix}
In the case of two-mode interferometers the function $\mathcal{D}^{M',M}_{k',k}(\hat{\Lambda}_{\mathbf{n}},\theta) = \langle M',k'|e^{i\theta \inter}|M,k\rangle$ can be expressed in terms of the Wigner rotation matrix $d^{(j)}_{m,m'}(\theta)$~\cite{WignerMatrix} for any values of $\theta$.
Therefore, when $\hat{\Lambda}_{\mathbf{n}}=\hat{\Lambda}^{(A)}_{\mathbf{n}} = \hat{Q}_{xy}$, we have
\begin{equation}
\mathcal{D}^{M',M}_{k',k}(\hat{\Lambda}^{(A)}_{\mathbf{n}},\theta) = i^{\frac{M-M'}{2}} \delta_{k'-\frac{M'}{2}, k-\frac{M}{2}}  
\, d^{\left(k-\frac{M}{2}\right)}_{\frac{M'}{2},\frac{M}{2}}(2\theta),
\end{equation}
and when $\hat{\Lambda}_{\mathbf{n}}=\hat{K}_y$,
\begin{align}
\mathcal{D}^{M',M}_{k',k}(\hat{K}_y,\theta)& =i^{\frac{3(k-k')}{2}-\frac{M-M'}{2}} \delta_{M'-k', M-k} \nonumber \\
{}& d^{\left(\frac{N+M-k}{2}\right)}_{\frac{3k'-N-M'}{2},\frac{3k-N-M}{2}}(2\theta).
\end{align}
Obviously $\lim\limits_{\theta \to 0}\mathcal{D}^{M',M}_{k',k}(\hat{\Lambda}_{\mathbf{n}},\theta) = \delta_{M,M'}\delta_{k,k'}$.

\bibliography{bibliography.bib}
\end{document}